\documentclass[structabstract]{aa}  
\usepackage{graphicx}

\usepackage{natbib}
\usepackage{amsmath}
\usepackage{rotating}
\usepackage{lscape}
\usepackage{color}
\usepackage{latexsym}
\usepackage{amsfonts}
\usepackage{multirow}
\usepackage{rotating}
\usepackage{graphicx}
\usepackage{epsfig}
\usepackage{amssymb,amsbsy}
\bibpunct{(}{)}{;}{a}{}{,}
\newcommand{\ie}{$i.e.,\;$}
\newcommand{\eg}{$e.g.,\;$}

\newcommand{\cf}{$cf.,\;$}

\usepackage{txfonts}
\begin{document}
   \title{X-ray spectral properties of Seyfert galaxies and the unification scheme}


   \author{Veeresh Singh\inst{1,2}\fnmsep\thanks{veeresh@iiap.res.in}, Prajval Shastri\inst{1}
          \and Guido Risaliti\inst{3,4} }
   \institute{Indian Institute of Astrophysics, Bangalore 560034, India
           \and
              Department of Physics, University of Calicut, Calicut 673635, India 
            \and
              INAF-Osservatorio di Arcetri, Largo E. Fermi 5, I-50125 Firenze, Italy 
         \and
             Harvard-Smithsonian Center for Astrophysics, 60 Garden St. Cambridge, MA 02138, USA
             }

   \date{Received xxxx xx, xxxx; accepted xxxx xx, xxxx}

 
  \abstract
{}
{The unification scheme of Seyfert galaxies predicts that the observed differences between type 1 and type 2 Seyfert galaxies are solely 
due to the differing orientations of the toroidal-shaped obscuring material around AGN. 
The observed X-ray spectra of Seyfert type 2s compared to type 1s are expected to be affected by higher absorbing column density due to the 
edge-on view of the obscuring torus. We study the 0.5 - 10 keV X-ray spectral properties of Seyfert type 1s and type 2s with the aim to test the predictions of Seyfert unification scheme in the X-ray regime.}
%
{We use an optically selected Seyfert sample in which type 1s and type 2s have matched distributions in the orientation independent parameters of AGN and host galaxy. 
}
%
{The 0.5 - 10 keV {{\em XMM-Newton}} pn X-ray spectra of Seyfert galaxies are in general best fitted with a model consists of an absorbed power-law, 
a narrow Gaussian fitted to the Fe K$\alpha$ emission line and an often seen soft excess component characterized by either 
a thermal plasma model with temperature kT $\sim$ 0.1 - 1.0 keV and/or a steep power-law. 
The 2.0 - 10 keV hard X-ray continuum emission in several Seyfert type 2s is reflection dominated and suggests the Compton-thick obscuration. 
Results on the statistical comparison of the distributions of the observed X-ray luminosities in 
the soft (0.5 - 2.0 keV) and hard (2.0 - 10.0 keV) bands, the X-ray absorbing column densities, the equivalent widths of Fe K$\alpha$ line and 
the flux ratios of hard X-ray to [OIII] $\lambda$5007$\mbox{\AA}$ for the 
two Seyfert subtypes are consistent with the obscuration and orientation based unification scheme.}
{}


   \maketitle
%

\section{Introduction}

Seyfert galaxies are classified mainly into two classes,
based on the presence (type~1) or absence (type~2) of broad emission lines in their optical spectra \citep{Ant93}. 
The unification scheme of Seyfert galaxies hypothesizes that Seyfert type~1s and type~2s constitute the same parent 
population and appear different solely due to the differing orientations of the AGN axis {\it w.r.t.} the line-of-sight. The 
scheme requires the presence of toroidal-shaped dusty molecular obscuring material around the AGN.  In 
the type~2s, this torus is inclined edge-on and blocks the direct view to the nuclear region, while in type~1s the 
observer's line-of-sight is away from the plane of the torus and the central region is visible \citep{Ant85,Ant93,Urry95}.
\par
Several investigations in the literature have yielded results consistent with the predictions of this scheme, 
such as, the presence of broad emission lines in the polarized optical and infrared spectra of many 
Seyfert 2s \citep{Moran2000}, the biconical structure of the narrow line region \citep{Wilson96}, 
the similarity of CO masses  of the two Seyfert subtypes \citep{Maiolino97}, 
and similar nuclear radio properties of both the subtypes \citep{Lal11}. Inconsistencies with the predictions 
of the scheme remain however,  {\eg}the absence of hidden Seyfert~1 nuclei in several Seyfert~2s \citep{Tran01,Tran03}, 
Seyfert~1s being preferentially hosted in galaxies of earlier Hubble type \citep{Malkan98}, 
the lack of X-ray absorption in some Seyfert~2s \citep{Panessa02},
and Seyfert~2s having a higher propensity for nuclear starbursts \citep{Buchanan06}. 
Moreover, it has been evident that the sample selection plays a crucial role in 
testing the predictions of Seyfert unification scheme \citep{Antonucci02}. \cite{Ho01} have discussed that the optical and UV 
selected samples are likely to have inherent biases against the obscured  sources. IR selected samples can be biased 
towards unusually dusty as well as sources which have higher level of nuclear star formation \citep{Ho01,Buchanan06}.
X-ray selected samples have also been used to examine the validity of Seyfert unification \citep{Awaki91,Smith96,Turner97a,Turner97b,Turner98,Bassani99}. 
However, the Seyfert samples selected from flux limited surveys are likely to have obscured type 2 Seyferts 
that are intrinsically more luminous than the selected type~1 counterparts. 
\cite{Maiolino98} and \cite{Risaliti99} have shown the 
increased number of heavily obscured type 2 sources using a sample based on 
[OIII] $\lambda$5007$\mbox{\AA}$ line luminosity, suggesting the generally inherent bias against less luminous and 
heavily obscured sources in X-ray selected samples.
Hard X-ray selected samples are suppose to be less biased but cannot be granted to be free from biases against 
heavily obscured Compton-thick and low luminosity AGNs \citep{Heckman05,Wang09}. 
Recent Seyfert samples based on {\em INTEGRAL} and {\em Swift}/BAT surveys preferentially contain relatively large number 
of high luminosity and less absorbed Seyferts \citep{Tueller08,Treister09,Beckmann09}, 
possibly due to less effective area which limits the sensitivity 
only to ($\sim$10$^{-11}$ erg s$^{-1}$ cm$^{-2}$) bright sources. \par
The quest of testing the validity and limitations of the Seyfert unification with more 
improved and well defined samples still continues. Recent studies by 
\cite{Cappi06,Dadina08,Beckmann09} used 
less biased optically and X-ray selected samples and reported the results broadly consistent with unification,
nonetheless, issues related to sample selection still remains.
Keeping the above sample selection arguments in mind we opted to use a sample in which the distributions for type 1 and 
type 2 Seyferts are matched in orientation independent parameters. Such sample selection 
mitigates the biases generally inherent 
in flux limited surveys and also ensures that the two Seyfert subtypes are intrinsically similar within the 
framework of Seyfert unification scheme \citep{Schmitt03a,Lal11}.
\\
%
In the standard picture of the unification scheme, AGN powered by accretion on to supermassive black hole and surrounding 
broad line region clouds are embedded in an obscuring torus \citep{Ant85,Ant93,Urry95}. 
The X-ray emission originating from the AGN carries the imprints of obscuring and reprocessing  material. 
Therefore Seyfert X-ray spectra are expected to differ depending upon whether the AGN is viewed 
directly ({\ie}type 1s) and through the obscuring torus ({\ie}type 2s). 
In Seyfert type 2s, the intrinsic X-ray emission is attenuated by the obscuring torus 
and the observed flux as well as the X-ray spectral shape depend on the optical depth of the obscuring 
torus \citep{Awaki91,Smith96,Turner97a,Bassani99,Cappi06}.
Since the photoelectric absorption cross-section decreases with the increase in photon energy, 
an increasing fraction of primary high energy photons are expected to transmit through the torus. 
Therefore, Seyfert type 1s and type 2s are expected to show systematically 
less difference for the observed hard X-ray luminosities and more difference for the soft X-ray (E $<$ 2.0 keV) luminosities.
Since in type 2s, AGN is viewed through the obscuring torus therefore type 2s are expected to show systematically higher absorbing column densities 
than type 1s.
The absorbing column density can be measured via photoelectric absorption component in 0.5 - 10 keV spectral fits as long as 
obscuration is Compton-thin (N$_{\rm H}$ $<$ 10$^{24}$ cm$^{-2}$).
If obscuration is Compton-thick (N$_{\rm H}$ $>$ 10$^{24}$ cm$^{-2}$), type 2s become faint even at higher X-ray energies 
since high energy photons are Compton-down scattered to soft photons which eventually get absorbed or scattered out of the line-of-sight \citep{Matt2000}.
The equivalent width (EW) of Fe K$\alpha$ line also increases with the absorbing column density as it is measured against much depressed continuum and 
for Compton-thick sources it increases to $\sim$ 1 keV \citep{Ghisellini94,Levenson06}.
The flux ratio of hard X-ray (2.0 - 10 keV) to [OIII] $\lambda$5007${\mbox {\AA}}$ line emission can also be used as a diagnostic parameter 
for the amount of obscuration and type 2s compared to type 1s are expected to show lower flux ratios of hard X-ray to [OIII] \citep{Bassani99,Cappi06}.
In this paper we test these predictions of the unification scheme by comparing the properties such as the observed 
soft and hard X-ray luminosities, the X-ray absorbing column densities, the EWs of Fe K$\alpha$ line and 
the flux ratios of X-ray to [OIII] $\lambda$5007${\mbox {\AA}}$ for 
the Seyfert type 1s and type 2s. In other words, using 0.5 - 10 keV {\em XMM-Newton} data we test whether X-ray properties conform with 
the optical classification of Seyfert type 1s and type 2s in the framework of the orientation and obscuration based unification scheme.
The important aspect of our study is the sample selection in which type 1s and type 2s are matched in 
orientation-independent parameters.
\par
A description of the sample, observations and data reduction is given in the section 2 and 3, section 4 describes 
the spectral modeling of X-ray spectra and the statistical comparisons of X-ray spectral properties of the 
two Seyfert subtypes have been discussed in section 5.
Notes on individual Seyfert galaxies are given in the appendix.
In this paper, we assume H$_{\rm 0}$~=~71~km$^{-1}$~Mpc$^{-1}$, $\Omega$$_{\rm m}$~=~0.27, and $\Omega$$_{\rm vac}$~=~0.73.\par

\section[]{The Sample}
%
%
%
%
We use the sample of \cite{Lal11} which is consist of 20 (10 type 1 and 10 type 2) optically selected Seyfert galaxies. 
This sample was formulated to study the nuclear radio properties in the framework of Seyfert unification scheme and 
in this paper we attempt to test the predictions of Seyfert unification scheme in the X-ray regime using the same sample. 
Although, the sample was formulated to study parsec-scale radio emission and is constrained by VLBI observing 
feasibility criteria, it is adequately qualified to test the predictions of unification scheme in 
the X-ray regime since it is based on the orientation-independent parameters.
In this sample Seyfert galaxies are defined as radio quiet ($\frac{\rm F_{5~GHz}}{\rm F_{B-band}}$ $<$ 10) \citep{Kellerman89}, 
low optical luminosity AGN (M$_{\rm B}$ $>$ -23) \citep{Schmidt83}, hosted 
in spiral or lenticular galaxies \citep{Weedman77}. All the intermediate subclasses Seyferts 
({\ie}1.0, 1.2, 1.5, 1.8, 1.9) which show any broad permitted emission line component in their optical spectra 
are grouped as type~1s, while those which show only 
narrow permitted emission lines are considered as type~2 Seyferts. 
%
The sample is selected such that the two Seyfert subtypes have matched distributions 
in the orientation-independent parameters of AGN and host galaxy and 
the sources which were deviating in matching the type 1 and type 2 distributions of the orientation-independent parameters were left out. 
In other words, the sources should lie within the same range 
of values for a given parameter to enter into the sample. Also, it was ensured that 
in a given bin of a parameter distribution type 1s do not outnumber the type 2s and vice-versa. 
Indeed, there is a possibility to increase the sample size following the same sample selection criteria, 
however, we would like to emphasize that the more important is the sample selection criteria and not 
the sample size to rigorously test the predictions of the unification scheme. 
Larger but heterogeneous and biased sample is likely to result incorrect conclusions.\par
Cosmological redshift, [OIII] $\lambda$5007$\mbox{\AA}$ line luminosity, 
Hubble stage of the host galaxy, absolute stellar magnitude of the host galaxy and absolute magnitude of the bulge are considered as 
the orientation-independent properties for the sample selection. 
These properties do not depend on the orientation of the obscuring torus, AGN axis and the host galaxy. 
And, indeed all these properties are intimately linked to the evolution of AGN as well as 
host galaxy.
We refer reader to see \cite{Lal11} for the detailed description on the sample selection and on the matching of the distributions of the orientation-independent parameters for the two Seyfert subtypes.
The matching of type 1s and type 2s in orientation-independent isotropic parameters ensures that 
we are not comparing entirely intrinsically different sources selected from different parts of the 
(luminosity, bulge mass, Hubble type, redshift) evolution function.

\begin{table*}
\centering
\begin{minipage}{140mm}
\caption{{\em XMM-Newton} EPIC pn observations log}
\begin{tabular} {@{}cccccccc@{}}
\hline
Source        & Obs.         &  Obs.   & Obs.      &  Net expo. & ct/s  & Obs. ID  & R$_{\rm S}$ \\ 
name          &  mode/filter &  date   & time (ks) & time (ks)      &         &        &   (arcsec)      \\ \hline
MCG+8-11-11   &  SW/Medium   &2004-04-10 & 38.45  &   25.24        & 14.93      & 0201930201  & 35.0      \\
MRK 1218      &  SW/Thin     &2005-04-09 & 13.87  &   6.55         & 0.53       & 0302260201  & 20.0    \\  
NGC 2639      &  FW/Thin     &2005-04-03 & 26.12  &   4.23         & 0.24       & 0301651101  & 16.0    \\
NGC 4151      &  SW/Medium   &2006-11-29 & 52.81  &   24.60        &  9.57      & 0402660201  & 40.0    \\
MRK 766       &  SW/Medium   &2005-05-24 & 95.51  &   60.12        &  4.92      & 0304030101  & 40.0    \\
MRK 231       &  FW/ Medium  &2001-06-07 & 22.34  &   17.30        &  0.10      & 0081340201  & 30.0    \\
ARK 564       &  SW/Medium   &2005-01-06 & 101.77 &   69.07        &  39.67      & 0206400101  & 35.0    \\
NGC 7469      &  SW/Medium   &2004-12-03 & 79.11  &   54.95        &  21.59    & 0207090201  & 37.5  \\
MRK 926       &  SW/Thin     &2000-12-01 & 11.76  &   7.19         &  18.33      & 0109130701  & 37.5  \\
MRK 530       &  SW/Thick    &2006-06-14 & 16.82  &   11.42        &  12.93      & 0305600601  & 20.0    \\
MRK 348       &  FW/Medium   &2002-07-18 & 49.50  &   26.54        &  1.96      & 0067540201  & 22.5  \\
MRK 1         &  FW/Thin     &2004-01-09 & 11.91  &   8.66         &  0.07      & 0200430301  & 17.5  \\
NGC 2273      &  FW/Medium   &2003-09-05 & 13.02  &   5.21         &  0.08      & 0140951001  & 20.0    \\
MRK 78        &  FW/Thin     &2006-03-11 & 16.12  &   4.83         &  0.09      & 0307001501  & 22.5  \\ 
NGC 7212      &  FW/Thin    &2004-05-20 & 14.22   &  10.50         &  0.10      & 0200430201   &  20.0   \\  
MRK 533       &  LW/Thick   &2004-06-02 & 10.42   &  7.51          &  0.12      & 0200660101   & 17.5  \\                 
NGC 7682      &  FW/Thin    &2005-05-27 & 20.21   &  14.27         & 0.04       & 0301150501   & 16.0    \\ \hline 
\end{tabular} 
Notes: SW: Small Window, FW: Full Window, LW: Large Window, R$_{\rm S}$: radius of the circular 
region around the source used for spectral extraction.  \par
\end{minipage}
\end{table*}

\section {X-ray observations and data reduction}
We modeled the X-ray spectra of our sample sources using 
{\em XMM-Newton} EPIC-pn observations obtained from {\em XMM-Newton} Science Archive. High throughput 
(especially, at energies E~$\geq$~2.0~keV) of {\em XMM-Newton} allows to search for spectral components with 
absorbing column density up to N$_{\rm H}$ $\sim$ 10$^{24}$ cm$^{-2}$ and its good spatial resolution 
minimizes any strong contamination from off-nuclear X-ray emission.
For our X-ray spectral study, we use only EPIC pn data and did not include 
EPIC MOS data since it is more susceptible to pile-up for bright X-ray sources 
and few sources, {\eg}MRK 530 do not have MOS data.
Also, EPIC pn data is good enough for our purpose of obtaining the average X-ray spectral parameters 
({\eg}photon index, equivalent hydrogen column density, EW of Fe K$\alpha$ emission line) 
of our sample sources. In case of a source had more than one observations in archive, we chose 
the latest one with sufficient long exposure time. The observation log is given in table 1 and gives the details 
of observing modes of pn camera, used optical light blocking filters, observation dates, observation and net exposure times, 
count rates, observation IDs and radius (in arcsec) of the circular region around the source position 
used to extract the X-ray spectra. 
The raw observation data files (ODFs) were reduced and analyzed using the standard Science Analysis 
System (SAS) software package (version 7.1) with the latest calibration files.
The time intervals of high energy background particle flares were removed by applying fixed threshold on single events, 
using E $\geq$ 10.0 keV, and $\delta$t = 50 sec light curves. The background is extracted from a 
circular region in the same chip. In order to apply $\chi^{2}$ statistics the 
spectra were binned with minimum 20 counts per bin. \par
We present the X-ray spectra of 17 (10 type 1s and 7 type 2s) out of 20 Seyfert galaxies of 
the sample using {\em XMM-Newton} pn archival data.
For 6 Seyfert galaxies, {\ie}MRK 1218, NGC 2639, MRK 530, MRK 78, MRK 533 and NGC 7682, the {\em XMM-Newton} X-ray spectra are presented for the first time and for two Seyferts, {\ie}NGC 4151 and NGC 7469, the {\em XMM-Newton} archival data chosen by us have not been published. The {\em XMM-Newton} spectra of rest of the 9 sources have been individually 
published, however, to have uniform reduction and analysis, we re-do the X-ray spectral fitting of these sources 
using data reduced with most updated calibration files. 
Also, a priori we do not rule out the possibility of obtaining the different or better fits for these 9 sources and therefore 
we prefer to re-do the spectral analysis for these sources. 
For 3 (type~2) Seyfert galaxies of our sample, {\ie}NGC 5135, MRK 477 and NGC 5929, there are no {\em XMM-Newton} data and 
we are using {\em Chandra} and {\em ASCA} spectral parameters for these sources from literature while performing 
the statistical comparison of the X-ray properties of the two subtypes of Seyferts.

\section{X-ray spectral analysis}
We attempt to obtain the physically motivated best fits for the 0.5 - 10.0 keV {\em XMM-Newton} pn X-ray spectra 
of our sample Seyfert galaxies.
Since the primary aim of spectral analysis is to identify the underlying continuum components, 
we first tried to fit each spectrum with a simplest model consists of a power-law 
plus absorption fixed at the Galactic value plus intrinsic absorption as a free parameter. 
However, in most of the cases this simple parametrization is insufficient to model the whole spectrum and 
residuals often show a soft excess and an emission line feature at 6.4~keV on top of the (absorbed or unabsorbed) power-law. 
In most of the cases, we found the best fit model consists of a power-law with a photoelectric cutoff due to 
cold absorption from the Galactic column density and neutral gas at 
the redshift of the source; a narrow Gaussian line to reproduce the iron K$\alpha$ fluorescent emission line 
at 6.4 keV in the rest frame
with energy, width and intensity as free parameters; soft excess component characterized by either a steep power-law and/or 
a thermal plasma model with temperature kT. In some cases reflection component characterized by 
reflection from an isotropically illuminated cold slab 
(model {\tt PEXRAV} \citep{Magdziarz95} in XSPEC \citep{Arnaud96}) is required for the best fit. 
A partial covering of the primary 
power-law component is also needed for the best fit in some sources. The partial covering model is equivalent to a 
combination of an absorbed power-law plus a scattered power-law, assuming that the power-law slopes of these two components are same. 
We note that for many of our sample sources, the 0.5 - 10.0 keV X-ray spectra can 
approximately be modeled as ``exp(-N$_{\rm H_0}$ $\sigma_{\rm abs}$) $\times$ 
[soft component + exp(-N$_{\rm H_1}$ $\sigma_{\rm abs}$) (AE$^{\Gamma}$ + emission lines)]'', as suggested by \cite{Nandra94,Awaki2000},
where N$_{\rm H_0}$ and N$_{\rm H_1}$ are the Galactic and source column densities, respectively. 
Table 2 describes the best fit spectral parameters of all the sample sources and 
figure 1 shows the best spectral fits of 17 sample sources.
Errors on the spectral fit parameters and derived quantities are at $90\%$ confidence level.\par
\subsection{X-ray spectral components}
In the following sections we describe the main components of the 0.5 - 10~keV spectra of our sample Seyfert galaxies.
\subsubsection{Absorbed power-law} 
The X-ray continuum of Seyfert galaxies can primarily be characterized by a photoelectric absorbed power-law. 
Figure 3 shows the distributions of equivalent hydrogen column density for the two subtypes of Seyferts of our sample. 
The estimated column densities in Seyfert galaxies of our sample varies from Galactic value to as high as 
the limiting value of Compton-thick obscuration (N$_{\rm H}$ $\ge$ 1.5$\times$ 10$^{24}$ cm$^{-2}$). 
As expected in the unification scheme, Seyfert type 1s of our sample 
are systematically less absorbed than type 2s. \par
Among type 2s, many sources ({\eg}MRK 1, NGC 2273, 
MRK 78, NGC 5135, NGC 7212, MRK 533 and NGC 7682) are heavily obscured and show only the reflection component 
in the 2.0 - 10 keV band. 
We identify these 7 type~2 Seyferts as Compton-thick sources with equivalent hydrogen column density along the line-of-sight 
1.5$\times$10$^{24}$ cm$^{-2}$ or higher and exhibiting reflection dominated spectra in the 2.0 - 10.0 keV band. 
The hard (2.0 - 10.0 keV) component of all the Seyfert type~1s is fitted with an absorbed power-law and 
our best fits render photon indices ($\Gamma$) $\sim$ 1.5 - 2.5 with the mean value of $\sim$ 2.0, while type 2s 
have either similar hard X-ray photon indices with high absorption or only pure reflection component.
In the cases, where both direct as well as reflection components are considered, we fix 
the reflection component photon index equal to that of the direct power-law component. 
In some cases equally good fits can be obtained either using unabsorbed or less absorbed 
flat ($\Gamma$ $\sim$ 0.5 - 1.0) power-law or with relatively more absorbed steep ($\Gamma$ $\sim$ 1.8 - 2.1) power-law. 
In such cases we preferred the more absorbed steep power-law fit 
as the power-law photon index is closer to the canonical value for Seyfert galaxies \citep{Cappi06}.
We do not perform the statistical comparison of the photon indices of the two Seyfert subtypes
since in many of the type~2 sources, the 2.0 - 10.0 keV hard X-ray spectrum is best fitted by pure reflection component and 
without the presence of any transmitted power-law component from AGN. Therefore in such reflection dominated sources we do not have
the photon index of the direct power-law component. 
\subsubsection{Soft X-ray excess} 
17 among the 20 Seyferts of our sample exhibit the soft excess, {\ie}emission above the extrapolation of the 
absorbed AGN power-law. 3 (type 1) sources, {\ie}MCG+8-11-11, MRK 1218 and MRK 926 do 
not show any presence of the soft excess. 
We fit the soft excess by a thermal plasma model {\tt MEKAL} (Mewe-Kaastra-Liedahl (1995)) and/or by a steep power-law.
The typical temperature of the thermal plasma corresponds to kT $\simeq$ 0.1 - 1.0 keV and the soft X-ray 
photon index is typically 2.5 - 3.5, however, in two cases (MRK 78, NGC 7682) the best fit photon index is 
as steep as $\sim$ 7.0 and 5.4, respectively.
In some cases ({\eg}NGC 4151, MRK 766 and NGC 7212), we see emission line features in 0.5 - 2.0 keV part of 
the spectrum, which indicates the presence of a sub-keV thermal plasma. \par
The origin of soft excess can be attributed to circumnuclear gas heated to million degrees by shocks induced by 
AGN outflows \citep{King05}, an intense circumnuclear star formation \citep{Cid98,Gonz01} or the photoionization and 
photoexcitation of circumnuclear gas by the primary AGN emission.
Recent high-resolution spectra of few sources, {\eg}NGC 1068 \citep{Young01,Kinkhabwala02}, 
the Circinus galaxy \citep{Sambruna01}, MRK 3 \citep{Sako2000,Bianchi05,Pounds05}; 
NGC 4151 \citep{Schurch04} have provided evidence that the soft excess is largely due to circumnuclear gas 
photoionized by primary AGN emission. 
Using a large sample of type 1 AGNs, \cite{Mateos10} noticed the presence soft X-ray excess in 
a substantial fraction of nearby sources and fitted it with a thermal plasma model with mean rest-frame 
temperature kT $\sim$ 100 eV. \cite{Mateos10} ruled out the origin of the soft excess as thermal emission from the accretion disk or Compton scattered disk emission on the basis of the temperatures detected and the lack of 
correlation of the soft excess temperature with the hard X-ray luminosity.

\par
\subsubsection{Reflection component}
Previous X-ray spectral studies reported that many Seyferts, in particular type~2s, 
show the presence of a reflection component \citep{Awaki2000}.
Compton reflection of nuclear radiation from optically thick material 
can be an important component in Seyfert type 2s and beside absorption, a flat spectrum can result from Compton 
reflection component \citep{Matt96}. 
In highly obscured (N$_{\rm H}$ $\geq$ 1.5 $\times$ 10$^{24}$~cm$^{-2}$), {\ie}Compton-thick sources, 
the primary radiation is almost completely absorbed (at E $\le$ 10 keV),
but these sources are still observable through the reflected radiation.
There are 7 type~2 sources in our sample in which the 2.0 - 10.0 keV part of the spectrum is best fitted with 
reflection component alone ({\tt PEXRAV} model in XSPEC; \cite{Magdziarz95}). 
The large EW of Fe K$\alpha$ line in these sources is consistent with the Compton-thick obscuration 
and reflection-dominated spectrum \citep{Ghisellini94,Levenson06}. 
Some type~1 sources, {\eg}MCG+8-11-11, NGC 4151, MRK 766 and MRK 926 also show the presence of a reflection component 
at the hard end of the spectrum. This can be expected 
in the framework of the unification scheme since a similar obscuring material having toroidal geometry is present but the
plane of the torus lies away from the line-of-sight and thus the reflection component is much fainter in 
compared to the direct transmitted component. 
Moreover, at the hard end of the spectrum (E $\geq$ 8.0 keV) the reflection component may have significant contribution.
\subsubsection{Fe K$_\alpha$ and other emission lines} 
We detected fluorescent Fe K$\alpha$ 
emission line at $\sim$ 6.4 keV in all (except NGC 2639) of our sample sources 
suggesting the ubiquitous presence of Fe K$\alpha$ line in the X-ray spectra of Seyfert galaxies. 
In most of the sources the centroid energy of Fe K$\alpha$ emission line is consistent with the emission from 
neutral or mildly ionized iron. 
In some sources, {\eg}MRK 766, MRK 231, ARK 564, MRK 533 and NGC 7682, an emission line at 
$\sim$ 6.7 - 7.1 keV is also present, which is possibly either H or He like ionized Fe K$\alpha$ line component or 
a neutral component of Fe K$_{\beta}$ emission line.
When the absorbing column density increases to a few times of 10$^{23}$ cm$^{-2}$, 
the equivalent width of Fe K$\alpha$ increases, since it is measured against a depressed continuum. 
The equivalent width can be higher than 1.0 keV for column densities of N$_{\rm H}$ $\geq$ 10$^{24}$ cm$^{-2}$, 
and such values are indeed observed in heavily absorbed and Compton-thick sources \citep{Maiolino98}.
Figure 3 shows the distributions of EWs of Fe K$\alpha$ line for the two subtypes of Seyferts of our sample. 
Seyfert type~1s have EWs $\sim$ 20 - 250 eV while for type~2s (except MRK 348) it ranges 
from $\sim$ 0.5 keV to $\sim$ 2.0 keV. High EWs of Fe K$\alpha$ line in type~2s suggests 
high obscuration compared to type~1s.\par
Apart from the ubiquitous Fe K$\alpha$ emission line we also notice the presence of other emission lines in some of our sample sources, 
mainly in the soft X-ray band.
The addition of Gaussian emission line at $\sim$ 0.88 keV in NGC 4151 and NGC 7212 improves the fit 
and is likely to be associated with O VIII radiative recombination continua \citep{Griffiths98}.
High resolution spectroscopic observations show that the soft X-ray emission in NGC 4151 is dominated by 
X-ray emission lines and radiative recombination continua (RRC) from hydrogen-like and helium-like ionization 
states of neon, oxygen, nitrogen and carbon \citep{Schurch04}. 
Also, in MRK 766 the addition of Gaussian emission feature at $\sim$ 0.5 keV improves the fit 
and may arise due to the blending of few narrow emission lines.
\section{Comparison of the X-ray spectral properties of the two Seyfert subtypes}
In this section we discuss the statistical comparison of the distributions of X-ray luminosities and spectral properties of 
the two subtypes of Seyferts in the framework of Seyfert unification scheme.  
\subsection{Soft (0.5 - 2.0 keV) and hard (2.0 - 10.0 keV) band X-ray luminosities}
Table 5 lists the fluxes, luminosities in the soft (0.5 - 2.0 keV) as well as hard (2.0 - 10.0 keV) X-ray bands and the 
flux ratios of the hard X-ray to [OIII] $\lambda$5007$\mbox {\AA}$ line emission.
Figure 2 shows the X-ray luminosity distributions for the two Seyfert subtypes in the soft (0.5 - 2.0 keV) as well as hard (2.0 - 10.0 keV) X-ray bands.
The observed soft X-ray luminosities for Seyfert type 1s in our sample range 
from $\sim$ 10$^{40}$ erg s$^{-1}$ to $\sim$ 10$^{44}$ erg s$^{-1}$ with 
the median value $\sim$ 6.58 $\times$ 10$^{42}$ erg s$^{-1}$, while for Seyfert type 2s it range from 
$\sim$ 10$^{38}$ erg s$^{-1}$ to $\sim$ 10$^{42}$ erg s$^{-1}$ with the median value $\sim$ 6.9 $\times$ 10$^{40}$ erg s$^{-1}$. 
The two sample Kolmogorov - Smirnov test shows that the soft (0.5 - 2.0 keV) X-ray luminosity distributions of type~1 and type~2 Seyferts 
are completely different, {\ie}null hypothesis that two distributions are same is rejected at 99$\%$ level of significance. 
Seyfert type~2s have systematically lower soft X-ray luminosities than type 1s. 
The nearly 2 Dex difference in the medians of the soft X-ray luminosities of two Seyfert subtypes can be explained as the presence of more 
line-of-sight obscuration in type 2s than that in type 1s, which is expected since torus intercepts observer's line-of-sight 
according to the orientation based Seyfert unification. 
In hard X-ray (2.0 - 10.0 keV) band, Seyfert 1s of our sample have luminosities ranging from 
$\sim$ 10$^{40}$ erg s$^{-1}$ to $\sim$ 10$^{45}$ erg s$^{-1}$ with the median value $\sim$ 1.1 $\times$ 10$^{43}$ erg s$^{-1}$, 
while for Seyfert 2s it range from $\sim$ 10$^{40}$ erg s$^{-1}$ to $\sim$ 10$^{44}$ erg s$^{-1}$ with the median 
value $\sim$ 6.8 $\times$ 10$^{41}$ erg s$^{-1}$. 
The difference in the hard X-ray luminosity distributions of the two Seyfert subtypes is 
lower compared to the soft band and can be explained as hard X-ray (2.0 - 10.0 keV) photons suffer less absorption.
According to the unification model, at sufficiently high energy one can expect 
similar X-ray luminosities for the two Seyfert subtypes as the torus optical thickness decreases with the increase 
in photon energy and 
similar central engines are at work in both subtypes. 
The substantial difference in the hard X-ray (2.0 - 10.0 keV) luminosity distributions of the two Seyfert subtypes can be understood as due to 
high absorption in type 2s. 
We have several heavily obscured (Compton-thick) type~2 sources in our sample 
in which X-ray continuum below 10.0 keV is completely suppressed and dominated by reprocessed components. 
\subsection{Line-of-sight obscuration}
The cold matter present in between observer and the AGN (consisting of 
the Galactic and the material at the redshift of AGN) preferentially absorbs soft X-ray from AGN. 
The photoelectric absorption component in the X-ray spectral modeling renders the amount of absorption along the line-of-sight 
in terms of the equivalent hydrogen column density, as long as direct transmitted component is seen. 
In general, we fit the X-ray continuum with a photoelectric absorbed power-law keeping column density and power-law photon index as free parameters 
and also all the spectral components are at least absorbed by the Galactic equivalent hydrogen column density. 
Figure  shows the distributions of cumulative N$_{\rm H}$ for the two Seyfert subtypes.
The estimated absorbing equivalent hydrogen column densities for our sample sources varies from as low as to 
Galactic value (N$_{\rm H}$ $\simeq$ 10$^{20}$ cm$^{-2}$) to as high as to the Compton-thick value 
(N$_{\rm H}$ $\geq$ 10$^{24}$ cm$^{-2}$). For type~1s the equivalent N$_{\rm H}$ ranges from $\sim$ 10$^{20}$ cm$^{-2}$ (Galactic value) to 
$\sim$ 10$^{22}$ cm$^{-2}$ with median value $\sim$ 10$^{21}$ cm$^{-2}$, while for type~2s it ranges from $\sim$ 10$^{22}$ cm$^{-2}$ to  $\geq$~10$^{24}$~cm$^{-2}$ with median value $\geq$~10$^{24}$~cm$^{-2}$. We have $\sim$ 
7/10 reflection dominated type~2 sources in which the 2.0 - 10.0 keV hard component is accounted by pure 
reflection component ({\tt PEXRAV} model) and therefore we do have only upper limit of N$_{\rm H}$ values ($\geq$ 1.5 $\times$ 10$^{24}$) in such cases. 
Seyfert type 2s have systematically higher absorbing column densities than types 1s and the distributions of 
the absorbing column densities (N$_{\rm H}$) are consistent with the orientation based unification scheme.
In our small sample we do not find any unobscured type~2 Seyfert, although some previous studies reported 
the presence of 10$\%$ - 30$\%$ of unobscured type~2 Seyferts \citep{Panessa02}.
While comparing the amount of obscuration for the two Seyfert subtypes we consider 
only cold absorption accounted by the photoelectric absorption component.
In several sources we require partial covering of primary AGN radiation. The best fit values of covering fraction and absorbing 
equivalent hydrogen column density are mentioned in table 3. 
In four Seyfert type 1s, {\ie}MRK 1218, NGC 4151, MRK 766 and NGC 7469, warm absorption characterized  
by the {\tt ABSORI} model (ionized absorber) is needed for the best fit. 
While using the {\tt ABSORI} model we fixed the photon index of the power-law of the absori component equal 
to the photon index of the primary power-law component and the temperature was fixed to 3$\times$10$^{4}$ K. 
The best fit parameter values for N$_{\rm H}$ and ionization parameters (${\rm \xi}$) are mentioned in table 4.
The ionization parameter indicates the ionization state of absorbing material and is defined as 
${\rm \xi}$ = $\frac{\rm L}{\rm nr^{2}}$, where L is the source luminosity (in $\sim$13.6 eV~-~13.6 keV), 
n is the absorbing gas density and r is the distance of absorber from the ionizing source (see; \cite{Osterbrock06}).
The estimated ionization parameter reveals low ionizing material in MRK 1218, NGC 4151 and MRK 766, however,
in NGC 7469, the ionization level is high. 
Using {\em XMM-Newton} RGS data, \cite{Blustin07} reported that 
NGC 7469 shows the evidence of ionized outflowing material with a wide range of ionization 
(log ${\xi}$ $\sim$ 0.5 - 3.5). 
\begin{landscape}
\begin{table}
\centering
\caption{The best fit spectral parameters} 
\small
\begin{tabular} {@{}ccccccccccc@{}}
\hline
Source & Model &      & \multicolumn{2}{c}{Soft X-ray Continuum} & \multicolumn{2}{c}{Hard X-ray Continuum} & \multicolumn{3}{c}{Emission Lines} & $\chi^2$/dof \\ 
      &     &N$_{\rm H}$  &$\Gamma_{\rm SX}$  & kT   & N$_{\rm H,hard}$ & $\Gamma_{\rm HX}$& E$_{\rm c,rest}$ & I$_{\rm line}$ & EW  &    \\
      &     &(10$^{20}$ cm$^{-2}$) &    & (keV) & (10$^{22}$ cm$^{-2}$) &   & (keV) &   & (eV)  &     \\ \hline
MCG+8-11-11& pha(PL+R+L)   &    18.30$_{-0.43}^{+0.40}$ &      &      &    & 1.84$_{-0.02}^{+0.02}$ & 6.44$_{-0.01}^{+0.01}$ &4.95 & 97.9$_{-18.5}^{+10.4}$  & 0.99 \\
MRK 1218 & pha(abs*PL+L)   & 15.20$_{-13.00}^{+4.30}$ &      &     &       & 1.48$_{-0.17}^{+0.11}$ & 6.32$_{-0.49}^{+0.22}$ &0.19 & 63.9$_{-52.4}^{+46.5}$  & 0.99 \\
NGC 2639 & pha(T+PL)       & 16.20$_{-15.60}^{+17.60}$ &      & 0.63$_{-0.09}^{+0.12}$ &       & 2.21$_{-1.02}^{+1.83}$ &     &     &       & 0.82 \\
NGC 4151 & pha(T+PL+pha*abs*pcf*PL+R+L+L) & 6.33$_{-2.60}^{+3.30}$ & 3.14$_{-0.06}^{+0.25}$ & 0.13$_{-0.01}^{+0.01}$ & 0.12$_{-0.03}^{+0.08}$ & 1.75$_{-0.03}^{+0.06}$ & 6.39$_{-0.01}^{+0.01}$ &14.43& 108.4$_{-6.7}^{+6.5}$ & 1.12 \\
         &                                   &      &      &      &       &     & 0.88$_{-0.01}^{+0.01}$ & 1.12& 60.1$_{-4.4}^{+29.0}$  &      \\
MRK 766  & pha(T+abs*pcf*PL+R+L+L+L)     &1.35$^{\rm f}$ &      & 0.20$_{-0.01}^{+0.01}$ &    & 2.17$_{-0.08}^{+0.10}$ &  6.44$_{-0.03}^{+0.03}$ & 0.50& 47.9$_{-13.5}^{+13.9}$  & 1.09 \\
         &                                   &      &      &      &       &      & 6.67$_{-0.07}^{+0.06}$ & 0.22& 19.3$_{-12.2}^{+12.6}$  &      \\
         &                                   &      &      &      &       &      & 0.48$_{-0.01}^{+0.01}$ & 5.38& 137.3$_{-37.5}^{+45.6}$ &      \\
MRK 231  & pha(T+T+pcf*pha*PL+L+L)       &1.25$^{\rm f}$ &      & 0.33$_{-0.03}^{+0.04}$ & 0.46$_{-0.16}^{+0.14}$  & 1.57$_{-0.30}^{+0.14}$ & 6.28$_{-0.13}^{+0.13}$ & 0.16& 215.0$_{-128.1}^{+123.1}$      & 1.04 \\
         &                                   &      &      & 0.99$_{-0.15}^{+0.14}$ &       &      & 6.66$_{-0.20}^{+0.07}$ & 0.15& 243.5$_{-163.5}^{+167.8}$  &      \\
ARK 564  & pha(T+PL+pcf*PL+L+L)           & 6.38$^{\rm f}$ & 3.42$_{-0.12}^{+0.13}$ & 0.97$_{-0.04}^{+0.04}$ &       & 2.32$_{-0.10}^{+0.08}$ &  6.33$_{-0.05}^{+0.05}$ & 0.29& 20.9$_{-9.9}^{+10.4}$ & 1.09 \\
         &                                   &      &      &      &       &      & 6.71$_{-0.03}^{+0.04}$ & 0.41& 32.3$_{-11.5}^{+11.4}$ &      \\
NGC 7469 & pha(T+abs*pcf*PL+L)          &4.96$^{\rm f}$ &      & 0.18$_{-0.01}^{+0.01}$ &     & 2.04$_{-0.02}^{+0.02}$ & 6.41$_{-0.02}^{+0.01}$ & 2.03& 64.2$_{-9.5}^{+7.5}$  & 1.09 \\
MRK 926  & pha(pcf*PL+R+L)                  &3.59$^{\rm f}$ &      &      &       & 1.98$_{-0.02}^{+0.03}$ & 6.33$_{-0.06}^{+0.06}$ &1.24 & 29.1$_{-23.8}^{+17.3}$  & 0.96 \\
MRK 530  & pha(T+pcf*PL+L)              &4.06$^{\rm f}$ &      & 0.20$_{-0.02}^{+0.02}$ &       & 2.28$_{-0.01}^{+0.03}$ & 6.40$_{-0.05}^{+0.05}$ & 1.49 & 47.2$_{-8.2}^{+7.4}$  & 1.09 \\
MRK 348  & pha(PL+pha*pcf*PL+L)            & 5.91$^{\rm f}$ & 2.75$_{-0.16}^{+0.17}$  &      & 6.86$_{-1.01}^{+0.84}$ & 1.70$_{-0.06}^{+0.07}$ &  6.40$_{-0.04}^{+0.04}$  & 1.59   &  34.2$_{-10.4}^{+10.8}$  &  1.05    \\
MRK 1    & pha(T+PL+R+L)          &5.31$^{\rm f}$ & 2.55$_{-0.29}^{+0.30}$ & 0.82$_{-0.10}^{+0.09}$ &$>$150 & 2.0$^{\rm f}$R& 6.77$_{-0.09}^{+0.08}$ &0.01 & 1249.2$_{-857.5}^{+1299.4}$& 0.99 \\
NGC 2273 & pha(PL+R+L)                      &7.75$_{-7.70}^{+36.20}$  & 2.79$_{-0.82}^{+1.87}$ &      &$>$150 &0.67$_{-0.92}^{+0.37}$R & 6.40$_{-0.02}^{+0.03}$ &2.74 & 2189.2$_{-441.7}^{+446.2}$& 0.97 \\
MRK 78   & pha(PL+R+L)              & 65.38$_{-24.48}^{+30.02}$ & 7.05$_{-1.84}^{+1.89}$ &      &$>$150 & 1.01$_{-0.54}^{+0.44}$R&  6.37$_{-0.09}^{+0.05}$ & 0.56& 673.0$_{-389.2}^{+402.9}$ & 0.85 \\
NGC 5135$^{\rm g}$ & pha(T+T+pha*PL+L)     & 4.58$^{\rm f}$ &      & 0.08 &$>$150& 1.5   &       6.40  & 0.52& 1700.0$_{-800.0}^{+600.0}$   &  ....     \\
            &                                &      &      & 0.39&       &      &     1.78  &  ....    &  ....    &       \\
MRK 477$^{\rm l}$ &     pha(pha*PL+L)        & 1.30$^{\rm f}$ &     &   & 24.0$_{-12.0}^{+17.0}$  & 1.9$^{\rm f}$ & 6.40$_{-0.23}^{+0.21}$  &  ....    & 560.0$_{-500.0}^{+560.0}$ &   0.35    \\
NGC 5929$^{\rm c}$ &    pha(PL+pha*PL+L)        & 51.60 & 1.7  &      & 27.7  & 1.7  &   6.40  &  ....    & 350.0&  ....     \\
NGC 7212 & pha(PL+R+L+L)                   & 5.41$^{\rm f}$ & 2.32$_{-0.20}^{+0.23}$ &      &$>$150 & 0.99$_{-0.37}^{+0.54}$R& 6.42$_{-0.04}^{+0.04}$ &     0.74 & 712.3$_{-133.4}^{+256.0}$ & 1.15  \\
         &                                   &      &      &      &       &      & 0.89$_{-0.04}^{+0.04}$ & 0.04 & 59.6$_{-38.5}^{+43.1}$ &      \\
MRK 533  & pha(T+PL+R+L+L)                & 5.16$^{\rm f}$ & 3.75$_{-0.61}^{+1.45}$ & 0.76$_{-0.10}^{+0.13}$ &$>$150 & 2.12$_{-0.51}^{+0.49}$R& 6.40$_{-0.05}^{+0.04}$ & 0.55  & 557.2$_{-292.7}^{+241.2}$& 0.97 \\
         &                                   &      &      &      &       &      & 7.04$_{-0.04}^{+0.06}$ & 0.53     & 665.6$_{-333.7}^{+387.9}$&    \\
NGC 7682 & pha(T+R+L+L)                   &34.88$_{-25.11}^{+57.82}$ & 5.36$_{-2.38}^{+2.78}$ &      &$>$150 & 1.77$_{-0.61}^{+0.58}$R& 6.43$_{-0.06}^{+0.18}$ & 0.20     & 477.5$_{-310.9}^{+312.3}$& 0.81 \\ 
         &                                   &      &      &      &       &      & 7.11$_{-0.09}^{+0.39}$ &0.13  & 457.6$_{-402.9}^{+460.3}$&    \\  \hline
\end{tabular}
\\
f: fixed; s: soft component; pha: photoelectric absorption; PL: power-law; 
L: Emission line fitted with a Gaussian; T: Thermal emission from hot gas ({\tt MEKAL} model in XSPEC); 
abs: Ionized absorber ({\tt ABSORI} model in XSPEC); R: Reflection from cold neutral material ({\tt PEXRAV} model in XSPEC); 
pcf: partial covering. \\ 
${\rm \bf g}$:~\cite{Guainazzi05a}; ${\rm \bf l}$:~\cite{Levenson01}; ${\rm \bf c}$:~\cite{Cardamone07}; 
emission line fluxes (I$_{\rm line}$) are in units of 10$^{-13}$ ergs cm$^{-2}$ s$^{-1}$.
\end{table}
\end{landscape}
%
\begin{table} 
\centering
\caption{Summary of partial covering model parameters}
\begin{tabular} {ccc}
\hline
Source      &  N$_{\rm H}$               &  Covering        \\ 
Name        &  (10$^{22}$ cm$^{-2}$) & fraction ({\it f})       \\ \hline
NGC 4151    &   9.06$_{-0.98}^{+1.27}$   &  0.69$_{-0.05}^{+0.04}$    \\
MRK 766     &   6.42$_{-0.56}^{+0.93}$   & 0.51$_{-0.03}^{+0.02}$   \\
MRK 231     &   8.35$_{-4.64}^{+4.02}$   & 0.74$_{-0.25}^{+0.09}$     \\
ARK 564     &   3.09$_{-0.45}^{+0.52}$   & 0.61$_{-0.15}^{+0.26}$     \\
NGC 7469    &  35.55$_{-6.77}^{+9.93}$ & 0.28$_{-0.03}^{+0.03}$   \\
MRK 926     &   4.35$_{-1.99}^{+2.86}$   & 0.15$_{-0.06}^{+0.06}$     \\
MRK 530     &   14.81$_{-4.18}^{+5.62}$  & 0.27$_{-0.27}^{+0.27}$     \\ 
MRK 348     &   10.47$_{-1.00}^{+0.68}$  & 0.84$_{-0.06}^{+0.06}$     \\ \hline
\end{tabular} 
\end{table}
\begin{table} 
\centering
\caption{Summary of ionized absorber parameters}
\begin{tabular} {ccc}
\hline
Source      &  N$_{\rm H}$               &  Ionization      \\ 
Name        &  (10$^{22}$ cm$^{-2}$) & parameter (${\rm \xi}$)       \\ \hline
MRK 1218    &   0.63$_{-0.32}^{+0.41}$   &  39.60$_{-38.37}^{+57.40}$    \\
NGC 4151    &   5.08$_{-1.20}^{+1.45}$   &  0.31$_{-0.28}^{+0.40}$    \\
MRK 766     &   0.21$_{-0.02}^{+0.03}$   & 6.61$_{-1.67}^{+4.67}$   \\
NGC 7469    &  1.06$_{-0.36}^{+0.42}$ & 1193.19$_{-232.63}^{+692.30}$   \\ \hline
\end{tabular} 
\end{table}
%
%
%
%
\begin{figure*}
\includegraphics[angle=270,width=8.0cm]{fig1-mcg8-11-11.eps}{\includegraphics[angle=270,width=8.0cm]{fig1-mrk1218.eps}}
\includegraphics[angle=270,width=8.0cm]{fig1-n2639.eps}{\includegraphics[angle=270,width=8.0cm]{fig1-n4151.eps}}
\includegraphics[angle=270,width=8.0cm]{fig1-mrk766.eps}{\includegraphics[angle=270,width=8.0cm]{fig1-mrk231.eps}}
\includegraphics[angle=270,width=8.0cm]{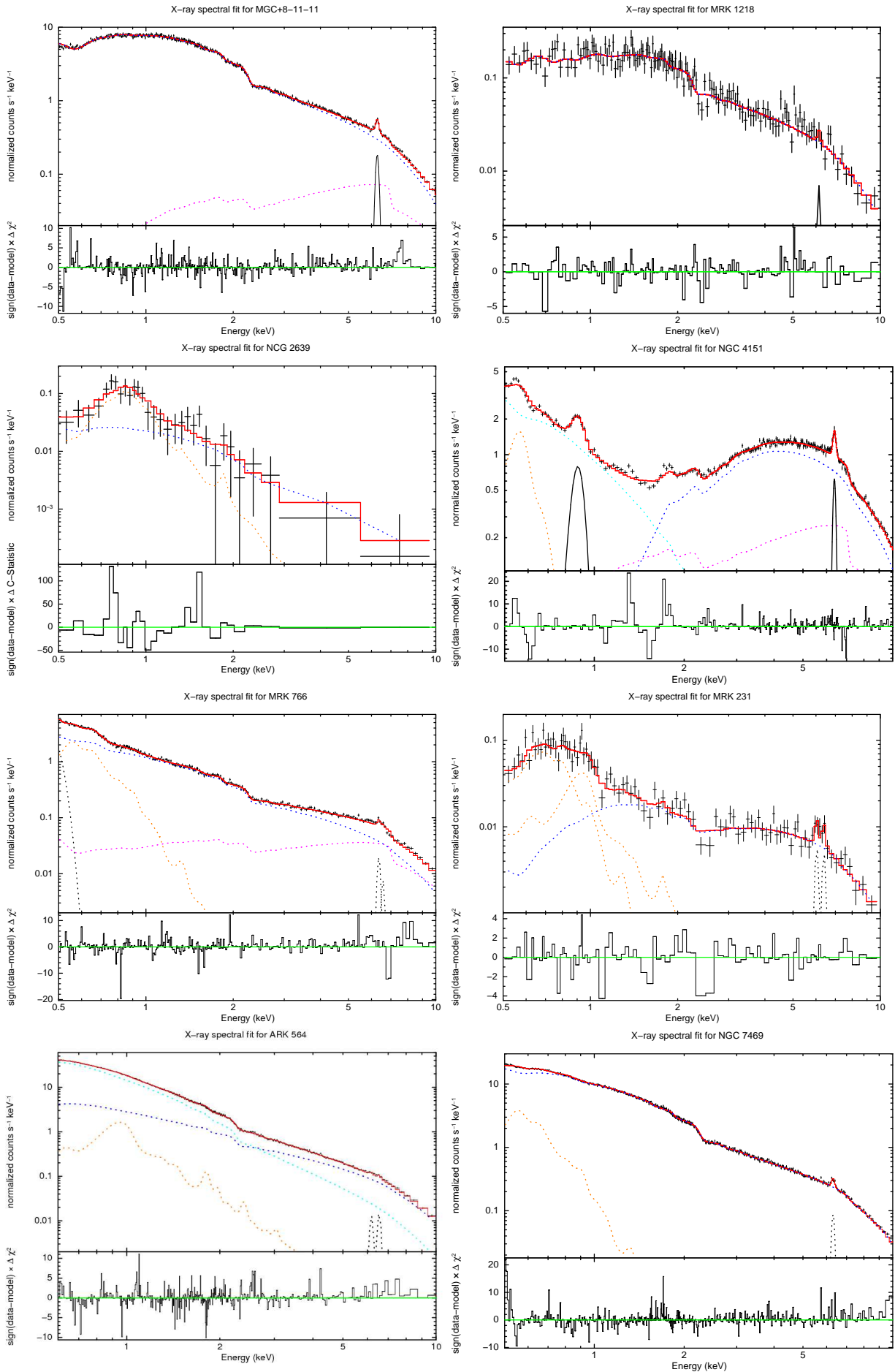}{\includegraphics[angle=270,width=8.0cm]{fig1-n7469.eps}}
\caption{ {\em XMM-Newton} pn 0.5 - 10.0 keV spectral fits for MCG+8-11-11, MRK 1218, NGC 2639, NGC 4151, MRK 766, MRK 231, ARK 564 and NGC 7469. For each source, top panel shows the cumulative fit (solid curve) along with all the additive spectral components (shown in dotted curves) against the spectral data points and bottom panel shows the residuals.}
\end{figure*}

\addtocounter{figure}{-1}

\begin{figure*}
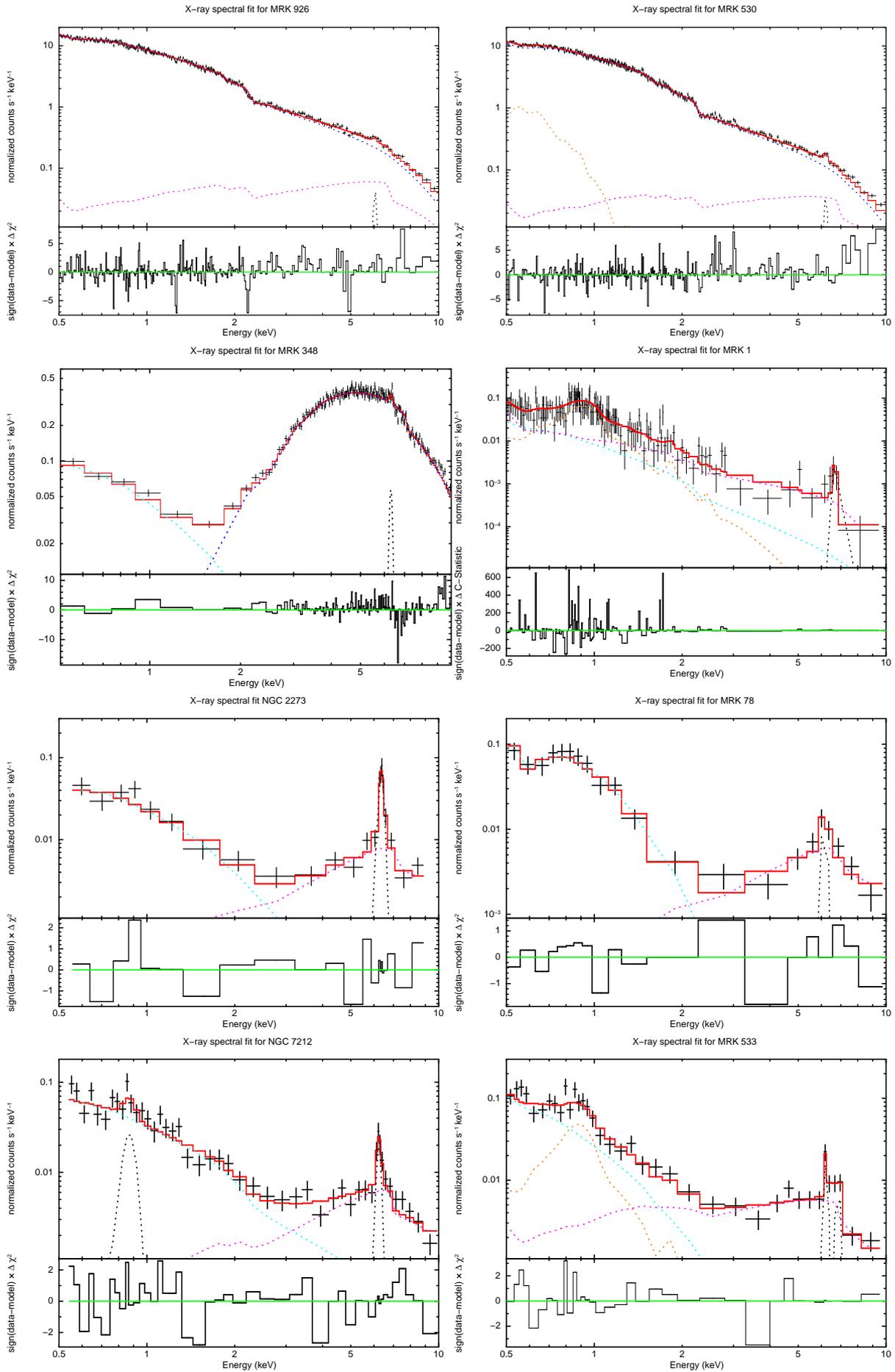

\includegraphics[angle=270,width=8.0cm]{fig1-mrk926.eps}{\includegraphics[angle=270,width=8.0cm]{fig1-mrk530.eps}}
\includegraphics[angle=270,width=8.0cm]{fig1-mrk348.eps}{\includegraphics[angle=270,width=8.0cm]{fig1-mrk1.eps}}
\includegraphics[angle=270,width=8.0cm]{fig1-n2273.eps}{\includegraphics[angle=270,width=8.0cm]{fig1-mrk78.eps}}
\includegraphics[angle=270,width=8.0cm]{fig1-n7212.eps}{\includegraphics[angle=270,width=8.0cm]{fig1-mrk533.eps}}
\caption{continued - spectral fits for MRK 926, MRK 530, MRK 348, MRK 1, NGC 2273, MRK 78, NGC 7212 and MRK 533.}
\end{figure*}
\addtocounter{figure}{-1}
\begin{figure*}
\includegraphics[angle=270,width=8.0cm]{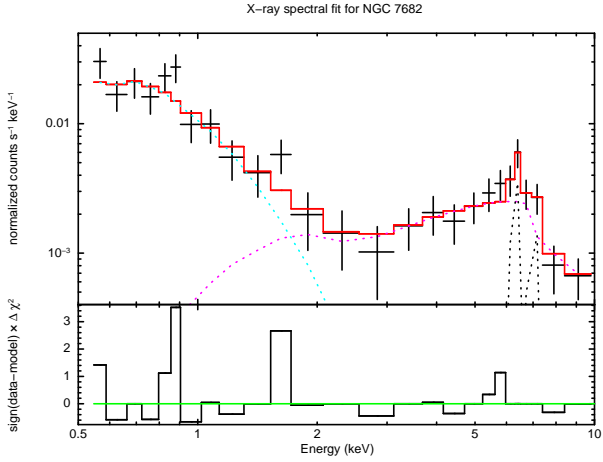}
\caption{continued - spectral fit for NGC 7682}
\end{figure*}
%
%
%
\begin{table*}
\begin{minipage}{140mm}
\caption{The observed X-ray fluxes and luminosities in the soft (0.5 - 2.0 keV) and hard (2.0 - 10.0 keV) bands and the flux 
ratios of hard X-ray to [OIII].}
\begin{tabular} {ccccccccccc} 
\hline
Name  & F$_{\rm 0.5 - 2.0~keV}$ & log L$_{\rm 0.5 - 2.0~keV}$  & F$_{\rm 2.0 - 10~keV}$ & log L$_{\rm 2.0 - 10~keV}$ & F$_{\rm [OIII]}^{\rm obs}$& Ref. & H$_{\alpha}$/H$_{\beta}$ & Ref. & F$_{\rm [OIII]}^{\rm cor}$ & logRx \\ \hline
MCG+8-11-11 & 161.18  & 43.35  & 441.97 & 43.79 & 6.43  & 1 &  4.37 & 7  & 19.43 & 1.36 \\
MRK 1218    &  4.86   & 41.96  & 25.35  & 42.68 & 1.70  & 2 & 3.26 & 8 & 2.17 & 1.07\\
NGC 2639    &  1.26   & 40.52  & 0.75  & 40.29  & 1.07  & 3 &  4.16 & 9 & 2.79 & -0.57  \\
NGC 4151    &  29.49  & 40.75  & 828.83 & 42.20 & 116.00 & 4 & 3.1  & 7 & 127.74 & 0.81  \\
MRK 766     &  33.99  & 42.10  & 72.66  & 42.43 & 3.95  & 1 & 5.1 & 7 & 18.79 & 0.59  \\
MRK 231     &  0.99   & 41.60  &  6.06  & 42.38 & 2.30  &2 & ...   & ... & ... & 0.42  \\
ARK 564     &  364.22 & 43.67  & 162.25 & 43.31 & 1.90  & 4 & ...  & ... & ... & 1.93 \\
NGC 7469    &  213.27 & 43.08  & 293.41 & 43.21 & 5.80  & 4 & 9.33  & 7 & 162.98 & 0.26 \\
MRK 926     &  171.95 & 43.94  & 299.05 & 44.18 & 3.50  & 4 & ...  & ... & ... &  1.93 \\
MRK 530     &  166.48 & 43.52  & 182.74 & 43.56 & 0.48  & 4 & ... & ... & ... & 2.58 \\
MRK 348     &  1.17   & 40.76  & 269.39 & 43.12 & 4.12  & 1 & 6.02 & 7  & 31.93 & 0.93 \\
MRK 1       & 0.98    & 40.74  & 0.81   & 40.66 & 6.00  & 4 & 5.89  & 7 & 43.61 & -1.73 \\
NGC 2273    &  0.53   & 39.62  & 10.02  & 40.89 & 1.60  & 5 & 6.92  & 7 & 18.68 & -0.27 \\
MRK 78      &  0.79   & 41.39  &  5.52  & 42.23 & 6.60  & 4 & 6.46   & 7 & 62.93 & -1.06 \\
NGC 5135$^{\rm g}$&1.90 & 40.91  & 1.60   & 40.83  & 3.70 & 6 & 7.8  & 9 & 61.66 & -1.59 \\
MRK 477$^{\rm l}$ &1.20 & 41.59  & 12.0   & 42.59  & 15.00 & 4 & 5.4  & 7 & 84.45 & -0.85 \\
NGC 5929$^{\rm c}$&0.81 & 40.15  &$>$14.0   & $>$41.39  & 0.93 & 4 & 5.5 & 7 & 5.53 & $>$0.40 \\
NGC 7212    & 0.83    & 41.13  &  6.96  & 42.05  & 8.75 & 1 & 5.01  & 7 & 39.52 & -0.75 \\
MRK 533     & 1.42    & 41.42  &  6.08  & 42.06  & 5.21 & 1 & 5.0  & 7 & 23.39  & -0.59 \\
NGC 7682    &  0.27   & 40.22  &  2.62  & 41.22  & 2.30 & 4  & 4.8  & 7 & 9.16 & -0.54 \\ \hline
\end{tabular}
\\
Notes: [OIII] and X-ray fluxes are in unit of 10$^{-13}$~erg~cm$^{-2}$~s$^{-1}$. 
R$_{\rm x}$: $\frac{\rm Flux (2.0 - 10.0~keV)}{\rm Flux [OIII]}$ ; 
${\rm \bf g}$: \cite{Guainazzi05a}; ${\rm \bf l}$: \cite{Levenson01}; ${\rm \bf c}$: \cite{Cardamone07}.
Ref., 1: \cite{Schmitt03a}, 2: \cite{Dahari88}, 3: \cite{Panessa06}, 4: \cite{Whittle92}, 5: \cite{Ferruit2000}, 6: \cite{Polletta96}, 7: \cite{Mulchaey94}, 8: \cite{Dahari88}, 9: \cite{Bassani99}.
\end{minipage}
\end{table*}
\begin{figure*}
\includegraphics[angle=0,width=6.3cm]{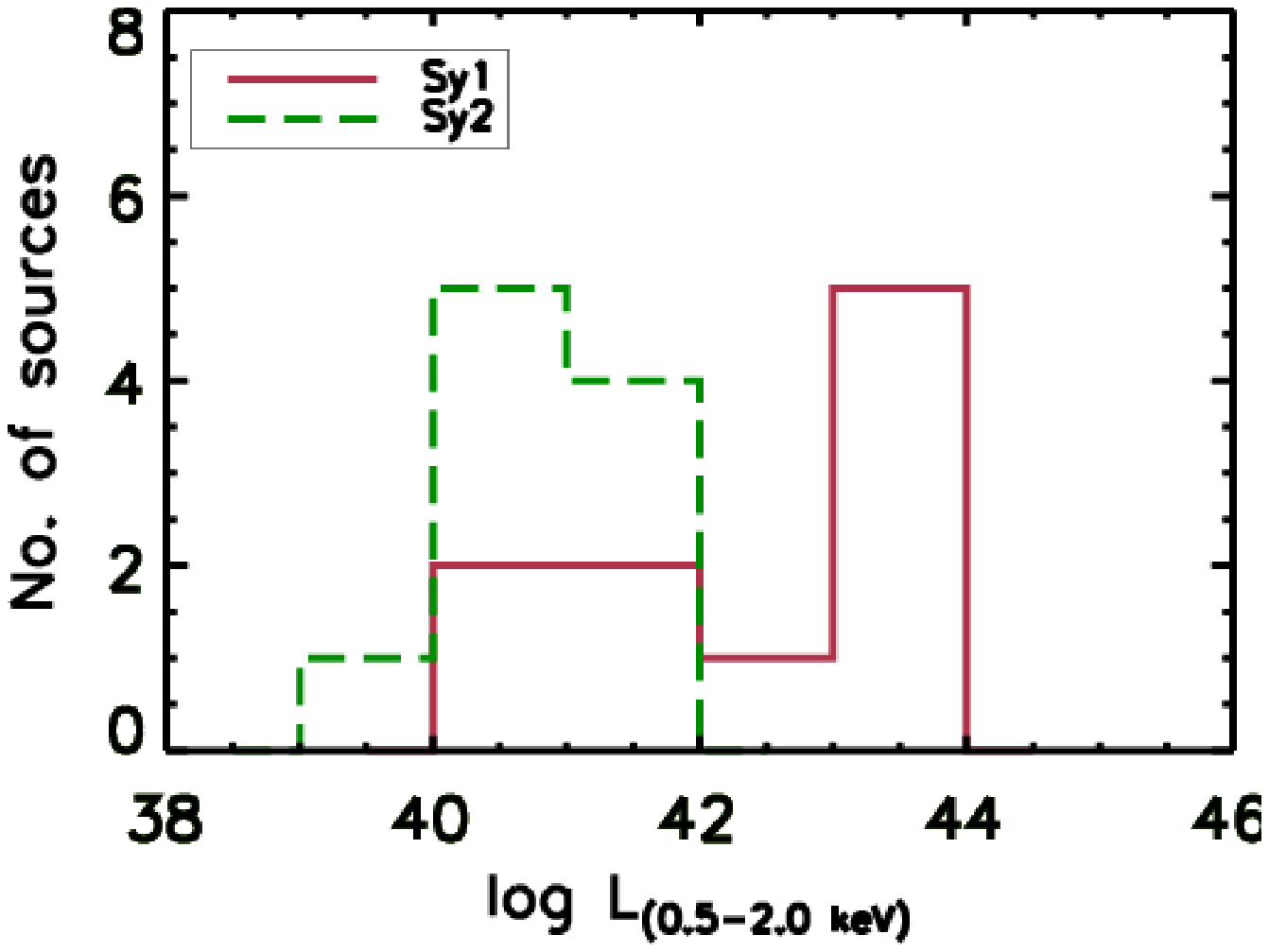}
\includegraphics[angle=0,width=6.3cm]{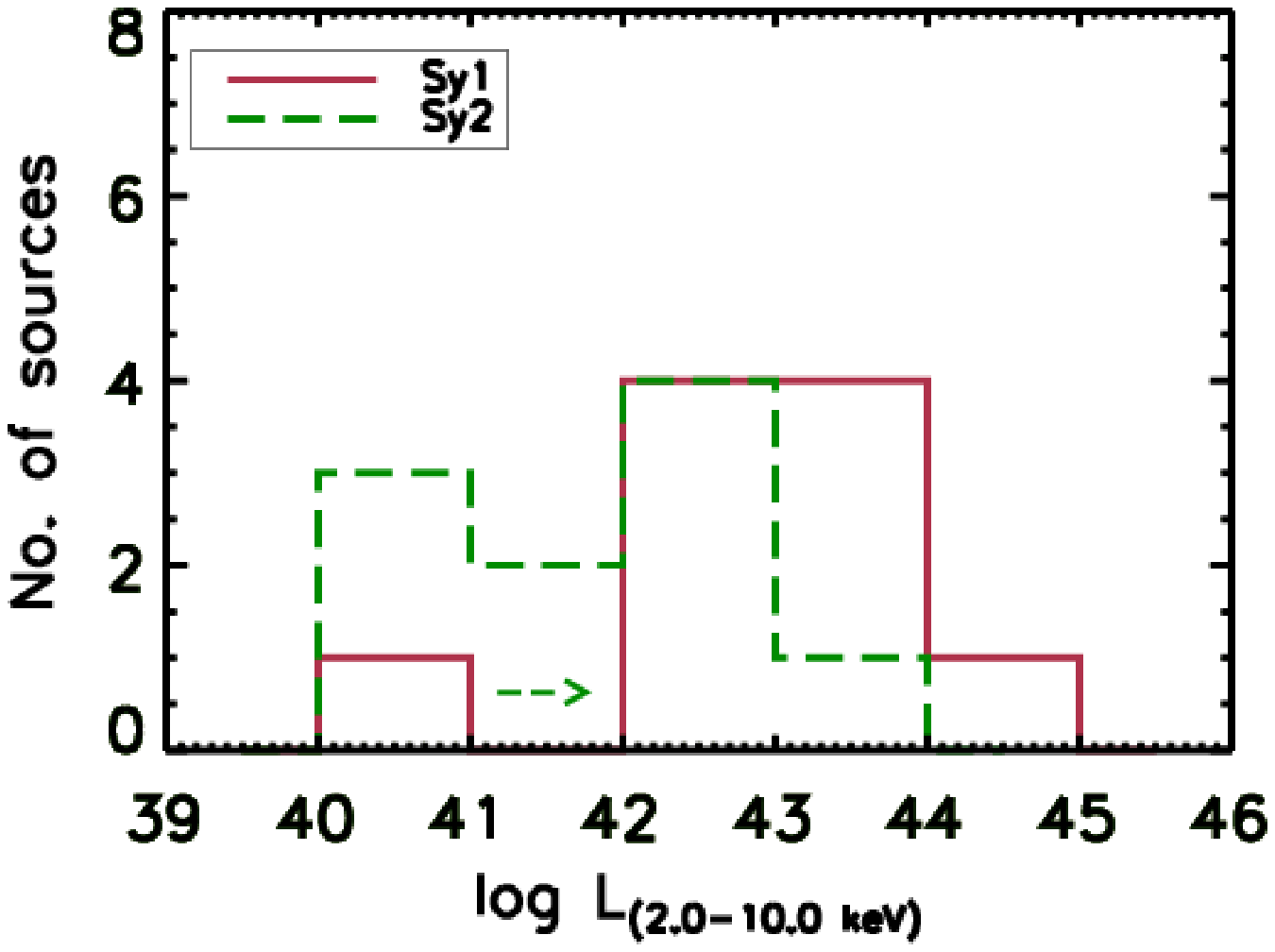}
\caption{Distributions of the soft X-ray (0.5 - 2.0) and hard X-ray (2.0 - 10.0) luminosities for the 
two subtypes of Seyfert galaxies.}
\end{figure*}
\begin{figure*}
\includegraphics[angle=0,width=6.3cm]{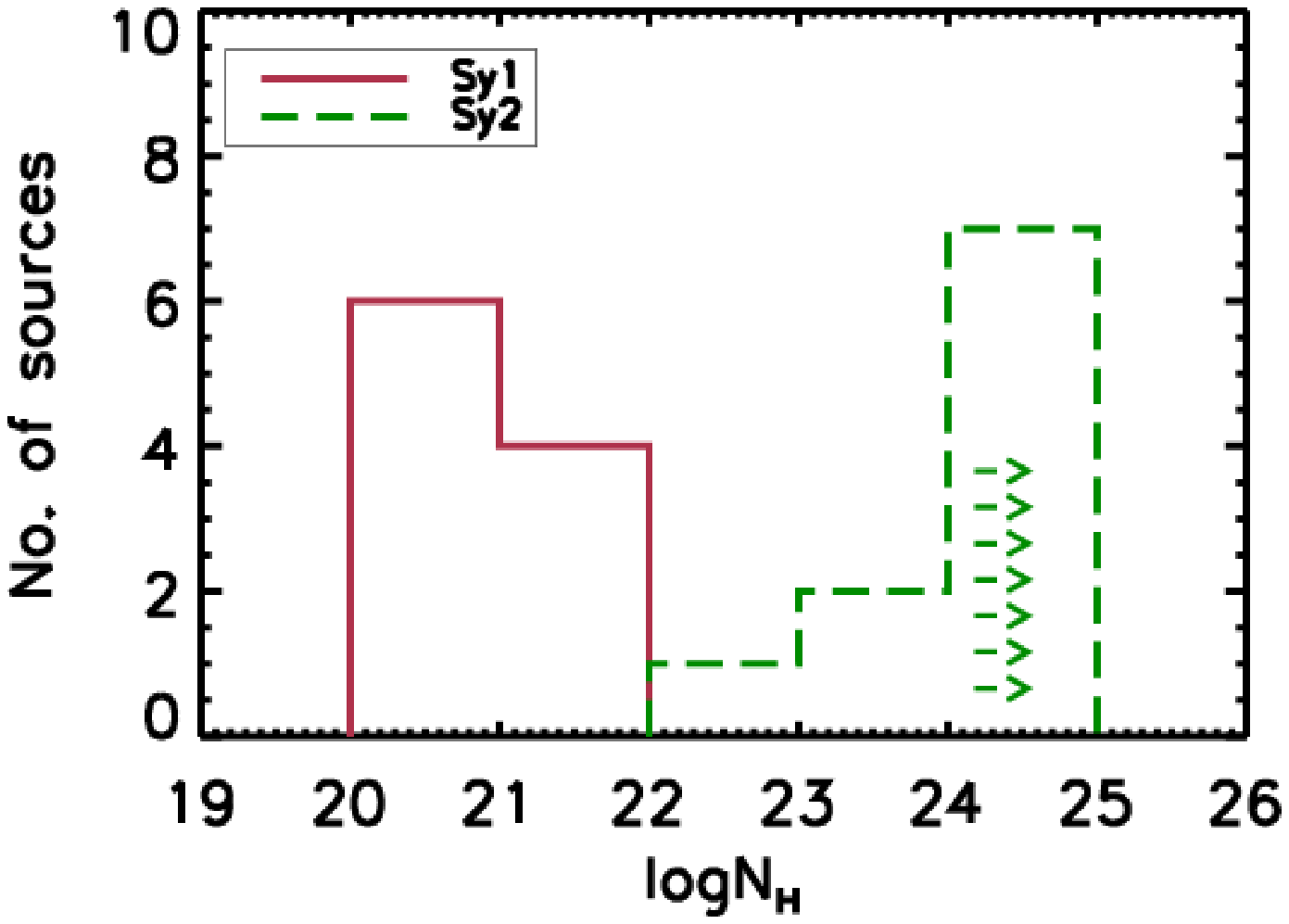}
\includegraphics[angle=0,width=6.3cm]{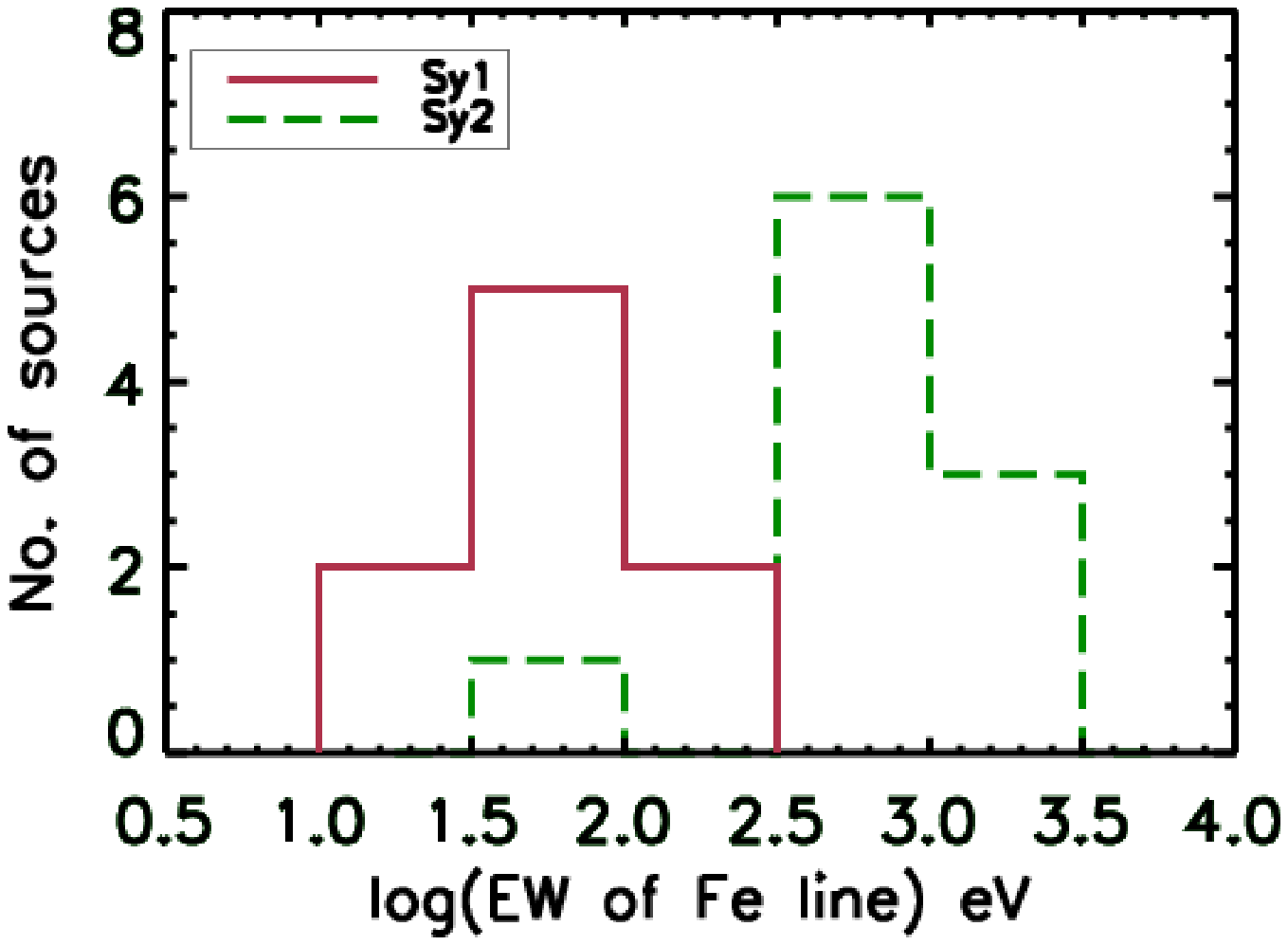}
\includegraphics[angle=0,width=6.0cm]{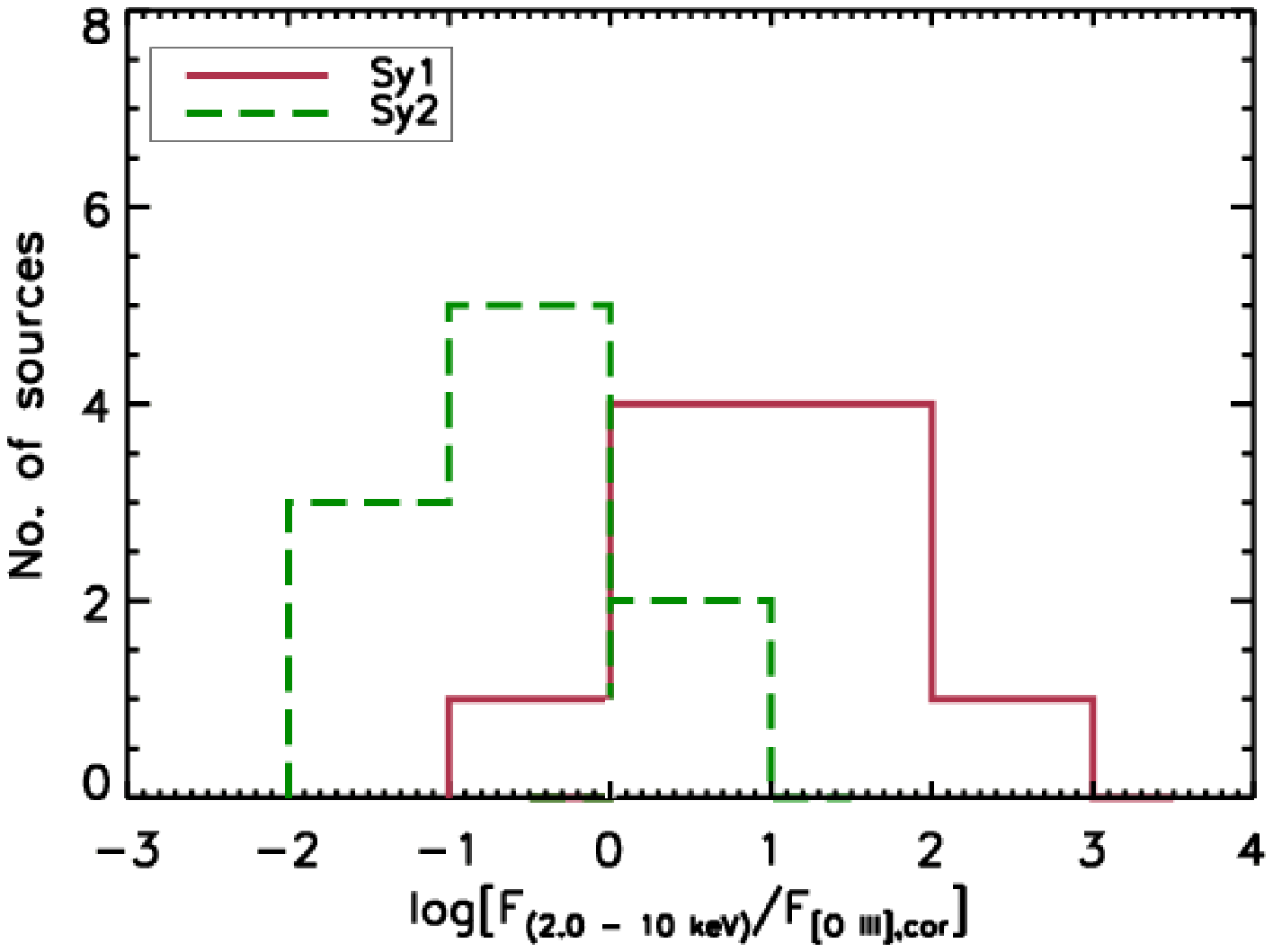}
\caption{Distributions of the equivalent hydrogen column density,
the equivalent width of Fe K$\alpha$ and the flux ratio of hard X-ray to [OIII] $\lambda$5007$\mbox{\AA}$
line for the two subtypes of Seyfert galaxies.}
\end{figure*}
\begin{table} 
\caption{Medians and Kolmogorov - Smirnov two sample tests for the statistical comparison of various distributions of the two Seyfert subtypes}
\begin{tabular} {ccccccc}
\hline
Distribution                 &    \multicolumn{2}{c}{Median}      & D       &  p-value    \\
                             & Type 1s & Type 2s      &                &                              \\ \hline 
log L$_{\rm 0.5 - 2.0~keV}$  &  42.82  & 40.84    & 0.8    &  2.06 $\times$ 10$^{-3}$           \\
log L$_{\rm 2.0 - 10.0~keV}$ &  43.02  & 41.83    & 0.6    &  5.25 $\times$ 10$^{-2}$          \\
log N$_{\rm H}$              &  21.02  &$>$ 24.18 & 1.0    &  9.08 $\times$ 10$^{-5}$            \\
log EW (eV)                  &  1.81   & 2.79    &   0.9    &  9.30 $\times$ 10$^{-4}$           \\ 
log R$_{\rm x}$              &  0.96   & -0.66   &  0.7     &  1.48 $\times$ 10$^{-2}$        \\ \hline
\end{tabular} 
\\
Kolmogorov - Smirnov two sample test examines the hypothesis that two samples comes from same distribution.
D = Sup x $|$S1(x) - S2(x)$|$ is the maximum difference between the cumulative distributions of two samples S1(x) and S2(x), respectively.
p-value is the probability that the null hypothesis, {\ie}two samples comes from same distribution, is correct.
R$_{\rm x}$ = $\frac{\rm Flux (2.0~-~10.0~keV)}{\rm Flux [OIII]}$. 
\end{table}
The presence of warm absorber in NGC 4151 and MRK 766 has also been reported in previous studies \citep{Schurch04,Mason03}.
%
\subsection{Equivalent width of Fe K$\alpha$ line}
We detect fluorescent Fe K$\alpha$ emission line in all but one (NGC 2639) of our sample sources suggesting the 
ubiquitous presence of Fe K$\alpha$ line in the X-ray spectra of Seyfert galaxies. 
In NGC 2639, due to lack of enough counts above 2.0 keV we could not confirm the presence of Fe K$\alpha$ emission line.
However, using {\em ASCA} observations of NGC 2639, \cite{Terashima02} reported the presence of Fe K$\alpha$ 
line with EW $\sim$ 1.5 - 3.0 keV.
We estimate the EW of Fe K$\alpha$ line with respect to the modeled continuum and find that Seyfert type~2s have systematically 
larger EWs ($\sim$ 350 eV to 2.2 keV, except of MRK 348), with median value $\sim$ 600 eV than that of type~1s ($\sim$ 20 - 200 eV) with 
median value $\sim$ 64.0 eV.
Kolmogorov - Smirnov two sample test shows that the distributions of EWs of Fe K$\alpha$ line of the 
two Seyfert subtypes are completely different ({\cf}table 6). 
Heavily obscured, {\ie}Compton-thick sources are characterized by large EW of Fe K$\alpha$ line 
\citep{Krolik94,Levenson06} and therefore we use the large EW ($\sim$ 1.0 keV) of 
Fe K$\alpha$ line as an indicator of heavy obscuration as it is measured against much 
depressed continuum (N$_{\rm H}$ $\sim$ 10$^{23}$ - 10$^{24}$ cm$^{-2}$; \cite{Leahy93}) or 
against pure reflection component (N$_{\rm H}$ $\sim$ 10$^{24}$ - 10$^{25}$; \cite{Bassani99}). 
However, identifying a source as the Compton-thick only on the basis of large EW of Fe K$\alpha$ line 
may not be correct, since a large EW can arise if the ionizing radiation is anisotropic \citep{Ghisellini91}, 
or if there is a lag between a drop in the continuum and line emissions, as observed in NGC 2992 by \cite{Weaver96}. 
\subsection{Flux ratio of hard X-ray to [OIII] $\lambda$5007$\mbox{\AA}$}
In order to confirm the high obscuration in Seyfert type~2s than in type~1s we use the flux ratio of hard (2.0 - 10.0 keV) X-ray to 
[OIII] $\lambda$5007$\mbox{\AA}$ and the equivalent width of Fe K$\alpha$ line as diagnostic tools. 
\cite{Bassani99} reported that the flux ratio of 
hard X-ray to [OIII] $\lambda$5007$\mbox{\AA}$ can effectively be used in identifying Compton-thin and Compton-thick sources 
with latter having flux ratio $\leq$ 1. The flux ratio of hard X-ray to [OIII] $\lambda$5007$\mbox{\AA}$ can be used as an indicator 
of the amount of obscuration since [OIII] originates from the NLR region largely 
unaffected by the torus obscuration and can be considered 
as the proxy of the intrinsic AGN power since it is produced by photoionization of NLR clouds by the AGN continuum \citep{Yee81,Nelson95}. 
While estimating the flux ratio of hard X-ray to [OIII] $\lambda$5007$\mbox{\AA}$, the latter ones are corrected for the optical reddening 
whenever Balmer decrement value was available for a source in literature. 
We have Balmer decrement (H$_{\alpha}$/H$_{\beta}$) values and hence, optical reddening corrected [OIII] fluxes for 
only 16 (6 type 1s and 10 type 2s) sources of our sample.
For reddening correction we use the formula given in \cite{Bassani99}, 
{\ie}F$_{\rm [OIII], cor}$ = F$_{\rm [OIII], obs}$[(H$_{\alpha}$/H$_{\beta}$)/${\rm (H_{\alpha}/H_{\beta}S)_{0}}$]$^{2.94}$, 
where intrinsic Balmer decrement ${\rm (H_{\alpha}/H_{\beta})_{0}}$ is assumed 3.0, and only narrow components of H$_{\alpha}$, H$_{\beta}$ 
are considered. \par
The statistical comparison of the distributions of the flux ratios of hard X-ray (2.0 - 10.0 keV) to [OIII] for the two 
Seyfert subtypes again confirms the systematically higher obscuration is type 2s than type 1s ({\cf}table 6, figure 3). 
Also, using this flux ratio as an indicator we note that MRK 1, MRK 78, NGC 5135, MRK 477, NGC 7212, MRK 533 and NGC 7682 have 
Compton-thick obscuration. 
X-ray spectral properties, {\eg}reflection dominated hard X-ray component, low hard X-ray to [OIII] flux ratio and 
high EW of Fe K$\alpha$ line implies that $\sim$ 7 - 8 out of 10 type 2 sources of our sample are Compton-thick 
(N$_{\rm H}$ $\geq$ 1.5 $\times$ 10$^{24}$ cm$^{-2}$). 
Some of our sample sources, {\eg}MRK 477 \citep{Bassani99}, NGC 5135, NGC 7212 \citep{Levenson06}, NGC 2273 \citep{Awaki09} 
and MRK 533 \citep{Bianchi05b} have already been reported as Compton-thick. 
In our sample we have only one Seyfert subtype 1.8 (MRK 1218) and  one subtype 1.9 (NGC 2639) 
which are included into type 1s and thus even if we include subtypes 1.8, 1.9 into type 2s, 
the relative fraction of Compton-thick sources remains high $\sim$ 60$\%$ (7-8 source out of 12 type 2s).
\section{Conclusions}
We investigated the 0.5 - 10 keV X-ray spectral properties of a sample of 20 Seyfert galaxies in which 
the two Seyfert subtypes have matched distributions in orientation independent parameters.
we summarize the conclusions of our study as below.
\begin{itemize}

\item X-ray spectra of our sample Seyfert galaxies, in general, are best fitted with a model consists of : 
a power-law with a photoelectric cutoff due to the cold absorption from the Galactic column density and 
neutral gas at the redshift of the AGN; a narrow Gaussian fitted to the Fe K$\alpha$ fluorescent 
emission line at 6.4 keV; often seen soft excess component characterized by either a steep power-law and/or 
a thermal plasma model with temperature kT $\sim$ 0.1 - 1.0 keV.

\item In several of the heavily obscured type 2 sources, X-ray spectra are completely dominated by reflection 
component and the hard part of spectra can be characterized with the {\tt PEXRAV} model and a prominent 
Fe K$\alpha$ line.

\item Seyfert type~1s have systematically higher observed soft X-ray luminosities than type~2s, as would be expected in the unification scheme with the hypothesis that in type~2s obscuring torus intercepts the AGN view 
and therefore absorbs the soft X-ray photons emanating from the AGN. The distributions of the observed hard (2.0 - 10 keV) 
X-ray luminosities of the two subtypes differ much less significantly, 
consistent with the prediction of unification, as optical thickness of obscuring material 
decreases with the increase in X-ray photons energy.

\item The X-ray absorbing column density for Seyfert type~1s in our sample ranges 
from $\sim$ 10$^{20}$ cm$^{-2}$ (Galactic value) to $\sim$ 10$^{22}$ cm$^{-2}$ with median value $\sim$ 10$^{21}$ cm$^{-2}$, 
while for type~2s it ranges from $\sim$ 10$^{22}$ cm$^{-2}$ to Compton-thick limiting value $\geq$~10$^{24}$~cm$^{-2}$ with median value 
$\geq$~10$^{24}$~cm$^{-2}$. The distribution of X-ray absorbing column densities (N$_{\rm H}$) for Seyfert galaxies of our sample 
is broadly consistent with the previous studies ({\eg}\cite{Cappi06,Akylas09}).

\item Fe K$\alpha$ fluorescent emission line has been detected in all (except one) of our sample sources 
suggesting the ubiquitous presence of Fe K$\alpha$ line in the X-ray spectra of Seyfert galaxies.  
The centroid energy $\sim$ 6.4 keV of the line is consistent with the emission from neutral or mildly ionized iron.
The equivalent width of Fe K$\alpha$ line in type 2s ranges $\sim$ 350 eV to 2.2 keV and is systematically 
higher than that of 1s ($\sim$ 20 - 200 eV).

\item  The statistical comparison of X-ray spectral properties, {\ie} the X-ray absorbing column densities, 
the hard X-ray spectral shape (absorbed power-law versus reflection dominated), {\bf the} EWs of Fe K$\alpha$ line, 
the flux ratios of hard X-ray to [OIII] $\lambda$5007$\mbox{\AA}$, and the luminosity distributions in soft 
and hard X-ray bands for the two subtypes of Seyfert galaxies are consistent with the orientation and obscuration based 
Seyfert unification scheme. 

\item We have argued for the importance of sample selection in testing the Seyfert unification scheme. And, using a sample based on the properties independent to the orientation of the obscuring torus, AGN and host galaxy, we show that the 0.5 - 10 keV X-ray spectral properties are consistent with the unification scheme.

\item  We also note a high fraction $\sim$ 70$\%$ of heavily absorbed likely to be Compton-thick sources among the type 2 Seyfert population. This high fraction is in agreement with the previous study reported by \cite{Risaliti99} based on a [OIII] $\lambda$5007$\mbox{\AA}$ luminosity selected sample of Seyfert 2s and implies that an unbiased sample is essential to estimate the accurate relative fraction of the heavily absorbed, {\ie}Compton-thick AGNs. 
Also, the heavily obscured AGNs may in part responsible for to account missing sources contributing to 
the Cosmic X-ray background \citep{Worsley05,Gilli07}. 
\end{itemize}

\begin{acknowledgements}
This work is based on observations obtained with {\em XMM-Newton}, an ESA science mission with instruments and contributions 
directly funded by ESA Member States and the USA (NASA). Also, this research has made use of the NASA/IPAC Extragalactic 
Database (NED) which is operated by the Jet Propulsion Laboratory, 
California Institute of Technology, under contract with the National Aeronautics and Space Administration. Authors 
thank to anonymous referee for useful comments and suggestions to improve the manuscript.

\end{acknowledgements}

\begin{appendix}

\section{Notes on individual sources}

In this section we describe the X-ray spectral properties of individual sources of our sample and the comparison of their X-ray 
spectral properties with the previous studies.

\subsection*{MCG+8-11-11}

MCG+8-11-11 has been observed by all the major X-ray satellites, with the exception of 
{\em Chandra}. The {\em ASCA} \citep{Grandi98} and
{\it BeppoSAX} \citep{Perola2000} data were well fitted by a fairly standard model
composed of a power-law, a warm absorber, a Compton reflection component, and an Fe~K$\alpha$ line.
\cite{Matt06} fitted the {\em XMM-Newton} EPIC pn spectrum with a model composed of an absorbed power-law 
(N$_{\rm H}$ $\simeq$ 1.83$^{+0.06}_{-0.03}$ $\times$ 10$^{21}$ cm$^{-2}$, $\Gamma$ $\simeq$ 1.805$\pm$0.015), a Compton reflection
component (with the inclination angle kept fixed to 30$^{\circ}$), a Gaussian
Fe K$\alpha$ line (EW $\sim$ 75$\pm${15}~eV), warm absorption (N$_{\rm H, warm}$ $\simeq$ 1.1$^{+0.06}_{-0.06}$ $\times$ 10$^{22}$ cm$^{-2}$) 
with the temperature of the material fixed at $10^6$~K and an absorption edge at 0.74~keV, corresponding to He-like oxygen.
Our best fit spectral parameters are broadly consistent with \cite{Matt06}, although we get a good fit without using warm absorption.
We also confirm the absence of a soft excess in this source using the {\em XMM-Newton} EPIC pn data. 

\subsection*{MRK 1218}
MRK 1218 does not have {\em Chandra} and {\em ASCA} observations but has been detected by RoSAT.
We present the {\em XMM-Newton} spectrum of MRK 1218 for the first time.  We find that 
an absorbed power-law (N$_{\rm H}$ $\simeq$ 1.52$^{+0.43}_{-1.30}$ $\times$ 10$^{21}$ cm$^{-2}$, $\Gamma$ $\simeq$ 1.48$^{+0.11}_{-0.17}$), 
with warm absorption (N$_{\rm H, warm}$ $\simeq$ 0.63$^{+0.41}_{-0.32}$ $\times$ 10$^{22}$ cm$^{-2}$)  
having plasma temperature fixed at 3~$\times$~10$^4$~K and an Fe~K$\alpha$~line of EW $\simeq$ 63.9 eV gives 
the best fit.

\subsection*{NGC 2639}

\cite{Terashima02} fitted the {\em ASCA} spectrum of NGC~2639 by a partially covered power-law 
(with N$_{\rm H,1}$ $\sim$ 0.08 ($\le$ 0.32) $\times$ 10$^{22}$~cm$^{-2}$ ,~N$_{\rm H,2}$ $\sim$ 32$^{+12.5}_{-30}$ $\times$ 10$^{22}$~cm$^{-2}$, 
~covering~fraction $\sim$ 0.89$^{+0.08}_{-0.81}$,~$\Gamma$~$\sim$~2.8$^{+1.0}_{-0.6}$~and~$\chi^2$/dof $\simeq$ 45.8/49).
They also attempted a model consisting of an absorbed power-law plus a Raymond-Smith (Raymond \& Smith 1977) 
thermal plasma modified by the Galactic absorption. 
The best fit parameters reported by \cite{Terashima02} using this model are 
N$_{\rm H,1}$ $\simeq$ 0.027 $\times$ 10$^{22}$~cm$^{-2}$~(fixed), ~kT $\simeq$ 0.80$^{+0.27}_{-0.40}$,
 ~abundances ~fixed ~to ~0.5 ~of ~solar, ~N$_{\rm H,2}$~$\sim$ 0.0 ($\le$ 0.31) $\times$~10$^{22}$~cm$^{-2}$,
 ~$\Gamma$ $\simeq$ 1.92$^{+0.70}_{-0.37}$~and~$\chi^2$/dof $\simeq$ 45.6/49. 
The reported equivalent widths of Fe~K$\alpha$ for these 
two models are 1.49$^{+11.11}_{-1.27}$~keV~and~3.13$^{+2.27}_{-2.00}$~keV, respectively.  
We present {\em XMM-Newton} X-ray spectrum of NGC~2639 first time and find that the best fit consists of an absorbed power-law 
(N$_{\rm H}$ $\simeq$ 1.62$^{+1.76}_{-1.56}$ $\times$ 10$^{21}$ cm$^{-2}$, $\Gamma$ $\simeq$ 2.21$^{+1.83}_{-1.02}$) plus a soft component modeled 
with a thermal plasma ({\tt MEKAL} in XSPEC) of temperature kT $\simeq$ 0.63$^{+0.12}_{-0.09}$. 
We do not detect Fe K$\alpha$ line in the {\em XMM-Newton} pn spectrum probably 
due to lack of sufficient counts above 2.0 keV.\par

\subsection*{NGC 4151}

NGC~4151 is one of the most extensively studied Seyfert galaxies in nearly all wavelengths. 
The {\em ASCA} and {\it BeppoSAX} X-ray spectra were fitted with a flat absorbed power-law 
($\Gamma$ $\simeq$ 1.65), a contribution from a cold reflector and a 
two-component absorber, an intrinsic neutral component with N$_{\rm H}$ $\simeq$ 3.4 $\times$ 10$^{22}$~cm$^{-2}$ and 
a highly ionized absorber with N$_{\rm H}$ $\simeq$ 2.4 $\times$ 10$^{23}$~cm$^{-2}$ \citep{Schurch02,Piro05}.
\cite{Yang01} reported the {\em Chandra} ACIS observation of NGC~4151 with notable extended soft X-ray emission 
on a scale of several hundred parsecs and a spatially unresolved hard X-ray ($\geq$ 2.0 keV) component.
The spectrum of the unresolved nuclear source is described by a heavily absorbed (N$_{\rm H}$ $\sim$ 10$^{22}$~cm$^{-2}$), 
hard power-law ($\Gamma$ $\simeq$ 0.3) plus soft emission from either a power-law ($\Gamma$ $\simeq$ 2.6) or 
a thermal (kT $\simeq$ 0.6~keV) component. 
We fitted the {\em XMM-Newton} pn spectrum with a model consisting of 
a soft component characterized with a power-law plus thermal plasma model with a temperature of kT~$\simeq$~0.13~keV 
and an emission line at 0.88~keV; a hard component characterized with an absorbed power-law 
(N$_{\rm H}$ $\simeq$ 1.8 $\times$ 10$^{21}$ cm$^{-2}$, $\Gamma$ $\simeq$ 1.75$^{+0.06}_{-0.03}$), 
with warm absorption (N$_{\rm H}$ $\simeq$ 5.2$^{+1.45}_{-1.20}$ $\times$ 10$^{22}$ cm$^{-2}$) and partial covering 
(N$_{\rm H}$ $\simeq$ 9.06$^{+1.27}_{-0.98}$ $\times$ 10$^{22}$ cm$^{-2}$, covering fraction {\it f} $\simeq$ 0.69$^{+0.04}_{-0.05}$) 
plus a reflection component and an Fe K$\alpha$ emission line at 6.4~keV. The Fe~K$\alpha$ line is fitted with a narrow Gaussian component with 
an EW~$\simeq$~108.4$^{+6.5}_{-6.7}$ eV.

\subsection*{MRK 766}
RoSAT and {\em ASCA} observations showed that the X-ray spectrum of MRK~766 can be described by a power-law of index 
$\sim$~1.6 - 2.0 (increasing strongly with flux), a reflection component 
and a narrow Fe~K$\alpha$ emission line (EW~$\sim$~100~eV) \citep{Leighly96}. 
Later observations with {\it BeppoSAX} found a steeper power-law ($\Gamma \simeq 2.2$) 
and evidence for an absorption edge at ~7.4 keV \citep{Matt2000a}.
Using {\em XMM-Newton} observations \cite{Miller06} reported the variations in the flux of Fe~K$\alpha$ line on short time scales (5 - 20~ks) 
and its strong correlation with the continuum emission. 
We fitted the {\em XMM-Newton} pn X-ray spectrum of MRK~766 with a less absorbed fairly steep power-law 
($\Gamma$ $\simeq$ 2.17$^{+0.10}_{-0.08}$, N$_{\rm H}$ fixed to Galactic value) with partial covering 
(N$_{\rm H}$ $\simeq$ 6.42$^{+0.93}_{-0.56}$ $\times$ 10$^{22}$ cm$^{-2}$, covering fraction {\it f} $\simeq$ 0.51$^{+0.02}_{-0.03}$) 
and warm absorption (N$_{\rm H}$ $\simeq$ 2.1$^{+0.28}_{-0.24}$ $\times$ 10$^{21}$ cm$^{-2}$), a reflection component at the hard end, 
a soft component fitted with a thermal plasma model of temperature kT~$\sim$~0.2~keV and narrow Gaussian emission lines at 
0.48~keV, 6.44~keV and 6.67~keV. The Fe~K$\alpha$ emission may have broad component but we get a better fit with a narrow emission line 
with centroid energy at 6.44~keV and a second narrow line at 6.67~keV. The latter line could be Fe~K$\alpha$ from highly ionized (Fe~XXV) 
material or a K$_\beta$ line from neutral matter. 

\subsection*{MRK 231}

X-ray observations by ROSAT, {\em ASCA} (Turner 1999; Iwasawa 1999; Maloney \& Reynolds 2000) and more recently
{\em Chandra} \citep{Gallagher02,Ptak03}, have revealed the presence of extended soft X-ray emission
of thermal origin that is likely to be associated with the circumnuclear starburst, 
and a hard and flat ($\Gamma$ $\simeq$ 0.7) power-law component as well as an Fe K$_{\alpha}$ emission line with EW $\sim$ 300 eV 
(the line was detected by {\em ASCA} but not by {\em Chandra}).
From combined {\em XMM-Newton} and {\it BeppoSAX} observations of MRK 231, \cite{Braito04} found a highly
absorbed (N$_{\rm H}$ $\sim$ 2 $\times$ 10$^{24}$ cm$^{-2}$) AGN component. 
We obtain the best fit with an absorbed power-law 
(N$_{\rm H}$ $\simeq$ 4.6$^{+1.4}_{-1.6}$ $\times$ 10$^{21}$ cm$^{-2}$, $\Gamma$ $\simeq$ 1.57$^{+0.14}_{-0.30}$) accompanied by a partial covering 
(N$_{\rm H}$ $\simeq$ 8.35$^{+4.02}_{-4.64}$ $\times$ 10$^{22}$ cm$^{-2}$, covering fraction {\it f} $\simeq$ 0.74$^{+0.09}_{-0.25}$), 
a soft component fitted with two thermal plasma models with temperatures kT $\simeq$ 0.33 keV and 0.99 keV, and narrow emission lines at 
6.28 keV and 6.66 keV. The soft X-ray component may have significant contribution from starburst activity \citep{Braito04}.

\subsection*{ARK 564}

\cite{Vignali04} analyzed two sets of {\em XMM-Newton} observations of ARK 564 taken a 
year apart (2000 June and 2001 June) and fitted 
the 0.6 - 10.0~keV continuum by a soft blackbody component (kT $\simeq$ 140 - 150~eV), 
a steep power-law ($\Gamma$~$\simeq$~2.50$-$2.55) and an absorption edge at a rest-frame energy of $\sim$~0.73~keV, corresponding to O~VII. 
No significant spectral changes were observed between the two observations, although 
the X-ray flux in the second observation is 40 - 50$\%$ lower. 
We obtain the best fit to the 0.6 - 10.0 keV {\em XMM-Newton} pn spectrum by a less absorbed steep power-law 
($\Gamma$ $\simeq$ 2.32$^{+0.08}_{-0.10}$, N$_{\rm H}$ fixed to the Galactic value) with partial covering 
(N$_{\rm H}$ $\simeq$ 3.09$^{+0.52}_{-0.45}$ $\times$ 10$^{22}$ cm$^{-2}$, covering fraction {\it f} 
$\simeq$ 0.61$^{+0.26}_{-0.15}$), a soft component fitted with a steep power-law 
($\Gamma$ $\simeq$ 3.42$^{+0.13}_{-0.12}$), a thermal plasma model with temperature kT $\simeq$ 0.97 keV and narrow emission lines at 6.33 keV and 6.71 keV. While fitting we excluded the 0.5 - 0.6~keV part of the spectrum 
since these data are of poor quality and make it difficult to obtain even a reasonably good fit.

\subsection*{NGC 7469}
\cite{Blustin03} fitted the 0.2 - 10.0~keV {\em XMM-Newton} EPIC pn spectrum with a combination of a Galactic absorbed power-law 
($\Gamma \simeq 1.75$), two blackbody models and a narrow Fe~K$\alpha$ emission line.
They noted significant residuals at low energies for which they suggested an instrumental origin.
We fitted the 0.5 - 10.0 {\em XMM-Newton} pn spectrum with a less absorbed power-law ($\Gamma$ $\simeq$ 2.04$^{+0.02}_{-0.02}$, N$_{\rm H}$ 
fixed to Galactic value), a partial covering (N$_{\rm H}$ $\simeq$ 35.55$^{+9.93}_{-6.77}$ $\times$ 10$^{22}$ cm$^{-2}$, 
covering fraction {\it f} $\simeq$ 0.28$^{+0.03}_{-0.03}$), a soft component fitted with a thermal plasma model of  
temperature kT $\simeq$ 0.18 keV and a narrow Gaussian fitted to the Fe K$\alpha$ line.
Using {\em ASCA} observations, \cite{Reynolds97} had reported  both narrow and broad components 
to the Fe line, which was supported by \cite{DeRosa02} using {\it BeppoSAX} data, 
and \cite{Nandra97a} suggested that it was relativistically broadened.
We fit the Fe~K$\alpha$ line with a narrow Gaussian component, however, and the data do not need 
a broad component for a good fit, consistent with \cite{Blustin03}.

\subsection*{MRK 926}
\cite{Bianchi04} fitted the combined 2.5 - 220~keV {\em XMM-Newton} and {\it BeppoSAX} spectrum with 
a baseline model consisting of a power-law ($\Gamma$ $\simeq$ 1.72$^{+0.08}_{-0.06}$) with intrinsic absorption 
over and above Galactic, a reflection component 
and a Gaussian to reproduce the Fe~K$\alpha$ line (EW $\simeq$ 45$^{+85}_{-24}$~eV). 
They also reported an upper limit to the Fe~K$_\beta $ flux of the order of the K$\alpha$ 
flux, and suggested the possibility of significant contribution from an Fe~XXVI line. \cite{Weaver01} 
reported strong line variability, both in centroid and in flux, between three {\em ASCA} observations.
We fitted the 0.5 - 10.0 keV {\em XMM-Newton} spectrum with an absorbed partially covered power-law 
($\Gamma$ $\simeq$ 1.98$^{+0.03}_{-0.02}$, N$_{\rm H}$ fixed to Galactic value, 
N$_{\rm part,H}$ $\simeq$ 4.35$^{+2.86}_{-1.99}$ $\times$ 10$^{22}$ cm$^{-2}$, covering fraction {\it f} $\simeq$ 0.15$^{+0.06}_{-0.06}$),
a reflection component and a narrow Gaussian to fit the Fe~K$\alpha$ line. 
Notably we do not see any soft excess in this Seyfert. \par

\subsection*{MRK 530}
MRK~530 has not been observed by either {\em Chandra} or {\em ASCA}.
We present {\em XMM-Newton} pn spectrum of MRK 530 for the first time. We fitted the 
0.5 - 10.0~keV spectrum by an  
absorbed partially covered power-law ($\Gamma$ $\simeq$ 2.28$^{+0.03}_{-0.01}$, N$_{\rm H}$ fixed to Galactic value, 
N$_{\rm part,H}$ $\simeq$ 14.81$^{+5.62}_{-4.18}$ $\times$ 10$^{22}$ cm$^{-2}$, covering fraction {\it f} $\simeq$ 0.27$^{+0.27}_{-0.27}$), 
a soft component modeled with thermal plasma at a temperature kT $\simeq$ 0.20 keV and a narrow Gaussian fitted to the Fe~K$\alpha$ line.

\subsection*{MRK 348}
\cite{Awaki06} fitted the {\em XMM-Newton} pn 0.2 - 10.0~keV spectrum of MRK~348 with an absorbed power-law 
(N$_{\rm H}$ $\simeq$ 13.5$\pm$0.02 $\times$ 10$^{22}$~cm$^{-2}$, $\Gamma$ $\simeq$ 1.61 $\pm$0.02), 
a soft component fitted with a thermal plasma model with temperature kT $\simeq$ 0.56$^{+0.01} _{-0.02}$~keV
and an Fe~K$\alpha$ line (EW $\sim$ 46.4$\pm$20~eV), and obtain $\chi^2$/dof $\sim$ 1.46.
They fixed the metal abundance of the thermal plasma at 0.1 times the solar abundance, 
the characteristic value of a normal galaxy \citep{Terashima02}. 
We use the same {\em XMM-Newton} data and find the best fit to the 0.5 - 10.0 keV spectrum consists of
an absorbed power-law (N$_{\rm H}$ $\simeq$ 6.86$^{+0.84}_{-1.01}$ $\times$ 10$^{22}$ cm$^{-2}$, $\Gamma$ $\simeq$ 1.70$^{+0.07}_{-0.06}$) 
with partial covering (N$_{\rm H, part}$ $\simeq$ 10.47$^{+0.68}_{-1.00}$ $\times$ 10$^{22}$ cm$^{-2}$, 
covering fraction {\it f} $\simeq$ 0.84$^{+0.06}_{-0.06}$), a soft component fitted with a power-law 
($\Gamma$ $\simeq$ 2.75$^{+0.17}_{-0.16}$) and a narrow Gaussian fitted to Fe K$\alpha$ line (EW $\sim$ 34.2$^{+10.8}_{-10.4}$) and obtained $\chi^2$/dof $\sim$ 1.05. \par

\subsection*{MRK 1}

\cite{Guainazzi05} fitted the 0.5 - 10.0~keV {\em XMM-Newton} pn spectrum of MRK 1 with 
a highly absorbed power-law (N$_{\rm H}$ $\ge$ 110 $\times$ 10$^{22}$~cm$^{-2}$, $\Gamma$ $\simeq$ 2.41$^{+0.13} _{-0.11}$) 
and a scattered component with partial covering factor fixed to 1, and obtained $\chi^2$/dof $\sim$ 36/32. They reported several emission lines including iron K shell 
lines in 0.5 - 2.0 keV band but no thermal component. 
In order to constrain the Fe~K$\alpha$ line they performed both global and local fits with 
centroid energy fixed at 6.4~keV, and found upper limits to the equivalent 
width of $\sim$~800~eV and $\sim$~2.0~keV respectively.
Due to the relatively low counts in the data, we use C-statistics to fit the spectrum and find the best 
fit consists of  a hard component (2.0 - 10.0~keV) completely dominated by reflection 
($\Gamma$ $\simeq$ 2.0), a soft component fitted with a power-law ($\Gamma$ $\simeq$ 2.55$^{+0.30} _{-0.29}$) and a thermal plasma model (kT $\simeq$~0.82$^{+0.08} _{-0.09}$~keV), 
and a Gaussian fitted to the Fe~K$\alpha$ line (EW $\sim$ 1.25$^{+1.29} _{-0.86}$ keV). The reflection 
dominated spectrum, high EW of the Fe line and the low ratio of hard X-ray to [OIII] flux suggest 
that MRK 1 is likely to be Compton-thick source.

\subsection*{NGC 2273}

\cite{Guainazzi05a} attempted to fit the {\em XMM-Newton} spectrum of NGC 2273 with 
a family of models in which hard X-ray component is  accounted by an absorbed power-law 
but yield an unacceptable flat intrinsic spectral index ($\Gamma$ $\simeq$ -0.2~-~-0.5) 
and also a large EW ($\simeq$ 2.0 - 3.6~keV) of the Fe~K$\alpha$ line with respect to the measured column density 
(N$_{\rm H}$ $\simeq$ 1.4 - 12.0 $\times$ 10$^{22}$~cm$^{-2}$), and suggested that the spectrum is dominated 
by Compton reflection. 
In the best fit reported by \cite{Guainazzi05a}, the soft X-ray spectrum is accounted by thermal emission model with temperature 
kT $\simeq$ 0.8$\pm$0.2 keV, hard X-ray fitted by reflection model ($\Gamma_{\rm hard}$ $\simeq$ 1.5$\pm$0.4,~N$_{\rm H}$ $\geq$ 18.0 $\times$ 10$^{23}$~cm$^{-2}$), and an Fe K$\alpha$ line with EW $\simeq$ 2.2$^{+0.4}_{-0.3}$ keV. 
We find the best fit in which hard component is completely accounted by reflection component ($\Gamma$ $\simeq$ 0.67$^{+0.37}_{-0.92}$ , 
{\tt PEXRAV} model), soft component is fitted with a power-law ($\Gamma$ $\simeq$ 2.79$^{+1.87}_{-0.82}$) and a Gaussian line fitted to 
Fe K$\alpha$ (EW $\sim$ 2.18$^{+0.45}_{-0.44}$ keV). 
Compton-thick nature of this source is evident from the reflection dominated spectrum, 
high EW of Fe K$\alpha$ line and low flux ratio of hard X-ray to [OIII].

\subsection*{MRK 78}
MRK 78 has not been observed by {\em Chandra} or {\em ASCA}.
\cite{Levenson01} fitted its {\it RoSAT} PSPC soft X-ray spectrum with a thermal plasma (kT $\simeq$ 0.76 keV) model 
absorbed by the Galactic column density and reported extended soft X-ray emission 
indicative of a circumnuclear starburst. We present the {\em XMM-Newton} spectrum of MRK 78 
for the first time. Our  best fit has the hard component completely accounted by 
a reflection component ($\Gamma$ $\simeq$ 1.01$^{+0.44}_{-0.54}$, `pexrav' model), the soft component fitted 
with a very steep power-law ($\Gamma$ $\simeq$ 7.05$^{+1.89}_{-1.84}$),  and a Gaussian fitted to Fe~K$\alpha$ 
(EW $\sim$ 0.67 keV). All the spectral components 
are absorbed by an equivalent hydrogen column density of N$_{\rm H}$ $\simeq$ 6.54$^{+3.0}_{-2.5}$ $\times$ 10$^{21}$~cm$^{-2}$. 
The predominance of the reflection component, high EW of the Fe~K$\alpha$ line and the ratio of hard X-ray flux 
to [OIII] line flux ratio suggest that obscuration is nearly Compton-thick.

\subsection*{NGC 5135}

NGC 5135 has not been observed by {\em XMM-Newton}, and we use the {\em Chandra} spectral 
properties given in \cite{Guainazzi05a} for our study. 
Both \cite{Guainazzi05a} and \cite{Levenson04} obtained substantially similar spectral fits consisting 
of thermal plus reflection model.
The soft X-ray spectrum requires two thermal components with kT $\simeq$ 80 eV and kT $\simeq$ 390 eV plus an additional emission line
with centroid energy E$_c$ $\simeq$ 1.78 keV. Above 2.0 keV the spectrum is Compton-reflection dominated, consistent with the AGN
being obscured by a column density N$_{\rm H}$ $\geq$ 9.0 $\times$ 10$^{23}$~cm$^{-2}$ (for an intrinsic photon index of 1.5 and a reflection
fraction $\leq$ 0.5). The EW of Fe K$\alpha$ fluorescent emission line against the reflection continuum is 1.7$^{+0.6}_{-0.8}$ keV. \\
Using {\em Chandra} observations \cite{Levenson04} reported a circumnuclear starburst in NGC 5135 and could 
spatially isolate the AGN emission which is entirely obscured by column density of N$_{\rm H}$ $\geq$~10$^{24}$~cm$^{-2}$, 
detectable in the {\em Chandra} bandpass only as a strongly reprocessed weak continuum represented by a flat power-law 
($\Gamma$ $\simeq$ 0) plus a prominent iron K$_\alpha$ emission line with EW of $\sim$ 2.4 keV. 
We confirm the Compton-thick obscuration in NGC 5135 using broad-band 0.5 - 50 keV {\em Suzaku} observations 
(Singh et al. 2010, in preparation).

\subsection*{MRK 477}

MRK 477 has not been observed by {\em XMM-Newton} or {\em Chandra},  and we therefore use 
the {\em ASCA} spectral properties given in \cite{Levenson01} for our study.
The best fit reported by \cite{Levenson01} consists of an absorbed power-law 
(N$_{\rm H}$ $\simeq$ 2.4$^{+1.7}_{-1.2}$ $\times$ 10$^{23}$~cm$^{-2}$,~$\Gamma$ $\simeq$ 1.9~fixed) and 
an unresolved Fe K$\alpha$ emission line of EW $\sim$ $560^{+560}_{-500}$~eV.  
The addition of a thermal component kT $\sim$ 0.9 keV to this model renders the reasonable spectral parameter values 
but did not statistically improve the fit.
\cite{Heckman97} pointed out that MRK~477 has a powerful circumnuclear starburst with the bolometric luminosity of 
$\sim$ 3 $\times$ 10$^{10}$ - 10$^{11}$ L$_{\odot}$. This luminosity is comparable to the AGN activity.  
However, \citet{Levenson01} concluded that the soft X-ray spectrum is dominated by the AGN. \par

\subsection*{NGC 5929}
NGC 5929 has not been observed by {\em XMM-Newton} or {\em Chandra}. 
\cite{Cardamone07} have modeled the 0.5 - 8.0~keV {\em ASCA} spectrum as a sum of three components: a weakly absorbed 
(N$_{\rm H,1}$ $\simeq$ 5.16 $\times$ 10$^{21}$~cm$^{-2}$) power-law with photon index $\Gamma$ $\simeq$ 1.70, 
a heavily absorbed power-law (N$_{\rm H,2}$ $\simeq$ 2.77 $\times$ 10$^{23}$~cm$^{-2}$) with the same photon index 
and a Gaussian Fe~K$\alpha$ line of EW of 0.35 keV centered at 6.19 keV. 
We use the {\em ASCA} spectral properties of NGC 5929 for our study.

\subsection*{NGC 7212}

\cite{Guainazzi05} fitted the 0.5 - 10.0~keV {\em XMM-Newton} spectrum with a heavily 
absorbed power-law (N$_{\rm H}$ $\ge$ 160 $\times$ 10$^{22}$~cm$^{-2}$, $\Gamma$ $\simeq$ 1.5$^{+0.3} _{-0.6}$), 
two thermal plasma components with kT $\simeq$ 0.16 and 0.72 keV and an Fe~K$\alpha$ line 
and obtain $\chi^2$/dof $\sim$ 97.0/65. The Fe K$\alpha$ line was fitted globally as well as locally 
giving EWs $\sim$ 900$^{+200}_{-300}$ eV and $\sim$ 1100$\pm$200 eV, respectively. 
We fitted the 0.5 - 10.0 keV {\em XMM-Newton} spectrum with the hard component completely accounted for by 
reflection ($\Gamma$ $\simeq$ 0.99$^{+0.54}_{-0.37}$), the soft component fitted with a power-law 
($\Gamma$ $\simeq$ 2.32$^{+0.23}_{-0.20}$),  a narrow Gaussian fitted to the emission line feature at 
$\sim$ 0.9 keV and  another Gaussian fitted to the 
Fe K$\alpha$ emission line at 6.4 keV giving EW $\sim$ 0.71$^{+0.25}_{-0.13}$ keV.

\subsection*{MRK 533}
MRK 533 has not been observed by  {\em Chandra} or {\em ASCA}. 
\cite{Levenson01} fitted its ROSAT/PSPC soft X-ray spectrum with a power-law $(\Gamma \simeq 2.0^{+0.7}_{-1.3})$. 
We present the {\em XMM-Newton} pn spectrum of MRK 533 for the first time.
We fitted the 0.5 - 10.0 keV spectrum with a model in which the soft component is fitted with a power-law 
($\Gamma$ $\simeq$ 3.75$^{+1.45}_{-0.61}$) and thermal plasma emission with the temperature kT $\simeq$ 0.76$^{+0.13}_{-0.10}$, 
a hard component completely accounted for by reflection ($\Gamma$ $\simeq$ 2.12$^{+0.49}_{-0.51}$, {\tt PEXRAV} model) and 
narrow Gaussians fitted to the Fe K$\alpha$ line  at 6.4 keV and the Fe K$_{\rm \beta}$ line at 7.04 keV, with 
EWs $\sim$ 0.56$^{+0.24}_{-0.29}$ keV and 
$\sim$ 0.67$^{+0.39}_{-0.33}$ keV, respectively.
Our X-ray spectral properties suggest the Compton-thick obscuration in MRK 533.

\subsection*{NGC 7682}
We present the {\em XMM-Newton} spectrum of NGC 7682 for the first time and find the best fit consists of 
a soft component fitted with a steep power-law ($\Gamma$ $\simeq$ 5.36$^{+2.78}_{-2.38}$), a  
hard component completely accounted for by reflection ($\Gamma$ $\simeq$ 1.77$^{+0.58}_{-0.61}$, {\tt PEXRAV} model), and
narrow Gaussians fitted to Fe K$\alpha$ line  at 6.4 keV and Fe K$_\beta$ line at 7.11 keV of 
EWs $\sim$ 0.48$^{+0.31}_{-0.31}$ keV and $\sim$ 0.46$^{+0.46}_{-0.40}$ keV, respectively. All the spectral 
components are absorbed by an equivalent hydrogen column density 
N$_{\rm H}$ $\simeq$ 3.48$^{+5.78}_{-2.51}$ $\times$ 10$^{21}$~cm$^{-2}$. Our {\em XMM-Newton} X-ray spectral properties 
suggest the Compton-thick obscuration in NGC 7682.

\end{appendix}

%
%
\bibliographystyle{aa}
\bibliography{VeereshSinghXMMpaper}

\end{document}